\documentclass{aa}  
\usepackage{natbib}
\bibpunct{(}{)}{;}{a}{}{,} 
\usepackage{graphicx}
\usepackage{txfonts}
\usepackage{lineno}
\usepackage{fixltx2e} 
\usepackage{longtable}
\usepackage{lscape}
\begin{document}
%

   \title{A global analysis of \textit{Spitzer} and new HARPS data confirms the loneliness and metal-richness of GJ\,436 b\thanks{Based on observations made with the HARPS spectrograph on the 3.6-m ESO telescope at the ESO La Silla Observatory, Chile.}
}

   \author{	A. A. Lanotte \inst{1},
              	M. Gillon \inst{1},
              	B.-O. Demory \inst{2,3},
         	J. J. Fortney \inst{4},
		N. Astudillo\inst{5}, 
              	X. Bonfils\inst{5},
          	P. Magain\inst{1},
		X. Delfosse\inst{5}, 
		T. Forveille\inst{5}, 
		C. Lovis\inst{6}, 
		M. Mayor\inst{6}, 
		V. Neves\inst{5,7,8}, 
		F. Pepe\inst{6}, 
		D. Queloz\inst{6,3},
		N. Santos\inst{8,9}, 
 		\and
		S. Udry\inst{6} 
          }

   \institute{Institut d'Astrophysique et de G\' eophysique (B\^{a}t. B5c), Universit\' e de
  Li\`ege, All\'ee du 6 Ao\^ut, 17, B-4000 Li\`ege, Belgium.\\
              \email{alanotte@ulg.ac.be}
         \and
         Department of Earth, Atmospheric and Planetary Sciences, Department of Physics, Massachusetts Institute of Technology, 77
Massachusetts Ave., Cambridge, MA 02139, USA
          \and
          Cavendish Laboratory, J J Thomson Avenue, Cambridge, CB3 0HE, UK
          \and          
         Department of Astronomy and Astrophysics, University of California, Santa Cruz, CA 95064, USA
        \and
UJF-Grenoble 1/CNRS-INSU, Institut de Plan\'etologie et d'Astrophysique de Grenoble (IPAG) UMR 5274, 38041 Grenoble, France
\and
Observatoire de Gen\`eve, 51 Ch. des Maillettes, 1290 Sauverny, Switzerland
\and
Departamento de F\'{\i}sica, Universidade Federal do Rio Grande do Norte, 59072-970 Natal, RN, Brazil
\and
Centro de Astrof\'{\i}sica, Universidade do Porto, Rua das Estrelas, 4150-762 Porto, Portugal
\and
Departamento de F\'{\i}sica e Astronomia, Faculdade de Ci\^encias, Universidade do Porto, Portugal
             }

   \date{Received ??, 2014; accepted ??, ??}

  \abstract
  {GJ\,436b is one of the few transiting warm Neptunes for which a detailed characterisation of the atmosphere is possible, whereas its non-negligible orbital eccentricity calls for further investigation. 
 Independent analyses of several individual datasets obtained with \textit{Spitzer}  have led to contradicting results attributed to the different techniques used to treat the instrumental effects.  }
{We aim at investigating these previous controversial results and developing our knowledge of the system based on the full \textit{Spitzer} photometry dataset combined with new Doppler measurements obtained with the HARPS spectrograph. We also want to search for additional planets.}
{  We optimise aperture photometry techniques and the photometric deconvolution algorithm DECPHOT to improve the data reduction of the \textit{Spitzer} photometry spanning wavelengths from 3-24 $\mu$m. Adding the high precision HARPS radial velocity data, we undertake a Bayesian global analysis of the system considering both instrumental and stellar effects on the flux variation.
  } 
  {We present a refined radius estimate of R$_P$=4.10 $\pm$ 0.16 R$_{\oplus}$ , mass M$_P$=25.4 $\pm$ 2.1 M$_{\oplus}$ and eccentricity $e$= 0.162 $\pm$ 0.004 for GJ\,436b. Our measured transit depths remain constant in time and wavelength, in disagreement with the results of previous studies. In addition, we find that the post-occultation flare-like structure at 3.6 $\mu$m that led to divergent results on the occultation depth measurement is spurious. 
  We obtain occultation depths at 3.6, 5.8, and 8.0~$\mu$m that are shallower than in previous works, in particular at 3.6~$\mu$m. However, these depths still appear consistent with a metal-rich atmosphere depleted in methane and
enhanced in CO/CO2, although perhaps less than previously thought. We could not detect a significant orbital modulation in the 8~$\mu$m phase curve. We find no evidence for a potential planetary companion, stellar activity, nor for a stellar spin-orbit misalignment. }
 {
 Recent theoretical models invoking high metallicity atmospheres for warm Neptunes are a reasonable match to our results, but we encourage new modelling efforts based on our revised data. 
  Future observations covering a wide wavelength range of GJ\,436b and other Neptune-class exoplanets will further illuminate their atmosphere properties, whilst future accurate radial velocity measurements might explain the eccentricity.
}


   \titlerunning{Global analysis of GJ\,436 \textit{Spitzer} and new HARPS data}
\authorrunning{Lanotte A. A. et al}
   \maketitle
%

\defcitealias{Knutson2011}{K11}
\defcitealias{Stevenson2010}{S10}
\defcitealias{Stevenson2012}{S12}
\defcitealias{Beaulieu2011}{B11}

\section{Introduction}

More than one thousand extrasolar planets have been discovered so far, most of them by the radial velocity (RV) and transit methods.
Transiting planets provide us with a wealth of information on their structure and atmospheric properties \citep[e.g.,][]{Winn2010b,SeagerDeming2010,Madhusudhan2014ppvi}. During the transit, the fraction of the stellar radiation transmitted through the planet's atmospheric limb can be measured  at different wavelengths  \citep{Charbonneau2002} to deduce the absorption spectrum of the planetary terminator. The occultation, i.e., the disappearance of the exoplanet behind its host star, enables the measurement of the dayside flux of the planet by spectroscopy and photometry \citep[e.g.,][]{Demingcommunity}. Because of signal-to-noise ratio (SNRs) limitations, most of the results in this field have been obtained by the photometric monitoring of multiple occultations in different broadband filters \citep[e.g.,][]{Charbonneau2005,Deming2005}. Occultation spectrophotometry also yields strong constraints on the planet's orbit \citep[e.g.,][]{Campo2011}. 

The bulk of the first measurements concerning exoplanetary atmospheres were gathered by the {\it Spitzer Space Telescope} \citep{Werner2004}. {\it Spitzer} was equipped with an 85-cm diameter telescope and three instruments to provide imaging and spectroscopic capabilities from 3.6 to 160 $\mu$m. These instruments are the Infrared Array Camera \citep[IRAC,][]{Fazio2004}, the Infrared Spectrograph \citep[IRS,][]{Houck2004}, and the Multiband Imaging Photometer \citep[MIPS,][]{Rieke2004}. {\it Spitzer} data are unique thanks to their large infrared wavelength range. It was fully operational from 2003 to 2009. It has been in the so-called  Warm mission since its cryogen was exhausted in May 2009, so that only two channels in the near-infrared (3.6 and 4.5 ${\mu}$m) are still operating. While ground-based facilities are now able to measure the thermal emission of some highly irradiated planets in the near-infrared \citep[e.g.,][]{Gillon2009corot1,Croll2011,Bean2013}, we have to wait for future space-based facilities like the {\it James Webb Space Telescope} \citep{Gardner2006} to complete these data at longer wavelengths.

Among the transiting extrasolar planets, GJ\,436b is one of the few low-mass planets orbiting a star which is small and nearby enough to allow an advanced characterisation. It is a $\sim$22 Earth mass planet discovered by radial velocity in orbit around a nearby ($\sim$10~pc) M2.5-type dwarf  \citep{Butler2004}. Its transit was originally detected by \cite{Gillon2007b} so that GJ\,436b became the first confirmed warm Neptune. {\it Spitzer} was then used to accurately measure its radius. Its low density suggests an envelope rich in H and He \citep{Gillon2007a,Deming2007}. 
This exoplanet is even more interesting because it shows a significant eccentricity that appears to disagree with tidal circularisation timescales and with the age of the system for an isolated planet \citep[e.g.][]{Demory2007}. Different authors tried to explain this with different mechanisms, usually with the presence of a companion \citep[e.g.,][]{Maness2007,Deming2007,Alonso2008,Ribas2008,Ribas2009,Caceres2009}. \cite{Beust2012} summarise those different explanations and propose an original solution based on a Kozai mechanism \citep{Kozai1962} assuming a distant disruptive body. More observations of the system should make it possible to test their hypothesis and constrain their model. In the meantime,  \cite{Stevenson2012} (hereafter S12) announced their possible detection of two new transiting Earth-sized companions to GJ\,436b by means of {\it Spitzer} data. 

Different surveys have been performed to characterise GJ\,436b's atmosphere. \cite{Pont2009} analysed two transmission spectra acquired with the Near Infrared Camera and Multi Object Spectrograph (NICMOS, 1.1-1.9 $\mu$m) on the Hubble Space Telescope (HST) in 2007.  Unfortunately, the strong NICMOS systematics led to inconclusive results. Recently, four transmission spectra were obtained by \cite{Knutson2014} with the Wide Field Camera 3 (WFC3) instrument on board the HST. They are featureless between 1.14 and 1.65~$\mu$m, ruling out cloud-free hydrogen-dominated atmosphere models. Besides, \citet{Kulow2014} showed through Lyman-$\alpha$ transit spectroscopy that GJ\,436b is probably trailed by a comet-like tail of neutral hydrogen. \citeauthor{Stevenson2010} (\citeyear{Stevenson2010}, hereafter S10) published their photometric observations ({\it Spitzer} program 40685) of GJ\,436b occultations in the 6 available bandpasses of the {\it Spitzer Space Telescope}, i.e. IRAC at 3.6, 4.5, 5.8, and 8~$\mu$m, IRS at 16~$\mu$m, and MIPS at 24 $\mu$m. They observed a high planetary flux at 3.6~$\mu$m, which they related to a possible depletion of methane in the atmosphere. In the meantime they could not detect the planetary emission at 4.5~$\mu$m, suggesting a high absorption coming most likely from CO and/or CO$_{2}$. This result was unexpected, as methane, and not CO/CO$_{2}$, should be the main carbon-bearing molecule in the relatively cold atmosphere of GJ\,436b ($T_{eq}=770$~K at periapsis assuming null albedo) according to the Gibbs free energy \citep{Burrows1999}.  These results were interpreted as the result of thermochemical disequilibrium by \cite{Madhu11a}.  
Shortly after, \citeauthor{Beaulieu2011} (2011, hereafter \citetalias{Beaulieu2011}) analysed  {\it Spitzer} observations of GJ\,436b transits obtained at 3.6, 4.5, and 8~$\mu$m. They measured a different planetary radius at 4.5~$\mu$m than at 3.6 and 8~$\mu$m, what made them conclude a high abundance of methane. In addition, \citetalias{Beaulieu2011} re-reduced the same occultation data as \citetalias{Stevenson2010} at 3.6 and 4.5~$\mu$m and obtained significantly different results.
\citetalias{Stevenson2010} and \citetalias{Beaulieu2011}  thus strongly differ in their interpretation, in particular on the IRAC 3.6 and 4.5 $\mu$m data during occultation. Both teams mentioned the need of a new dataset at those wavelengths to check their own theories. Later \citeauthor{Knutson2011} (2011, hereafter K11) invoked evidence of stellar variability to explain their (and \citetalias{Beaulieu2011}'s) planetary radius measurement discordance. They also improved the  system parameter estimates, notably thanks to a larger {\it Spitzer} dataset. 

It is not the first time that different teams have obtained conflicting results from the same {\it Spitzer} dataset. One can mention the detection \citep{Tinetti2007,Beaulieu2008} or non-detection \citep{Desert2009} of water vapour in HD\,189733b. 
In this context, we have decided to perform this new independent analysis of archived {\it Spitzer} data. The main concept of this project is not only to perform global Bayesian analysis of the extensive {\it Spitzer} datasets available for GJ\,436b and to compare them with the results previously obtained by different teams for subsets of the available data, but also to complement them with our new HARPS RV measurements to derive stronger constraints on the planetary properties. 
 When combined with the stellar properties, RV measurements provide orbital parameters and the minimal planetary mass. When we add in the transit light curves we can derive the mass and the radius of the planet, so that its mean density is known. Furthermore, the detection of an anomaly in the RV curve during the planetary transits, known as the Rossiter-McLaughlin effect, reveals the sky-projected angle between the stellar spin and planetary orbit axes \citep{Queloz2000}. The spin-orbit angle measurement may be helpful in modelling the planetary orbital evolution.
The High Accuracy Radial-velocity Planet Searcher \citep[HARPS--][]{Mayor2003}, installed in 2002 on a 3.6-m telescope at the La Silla Observatory (Chile), has demonstrated a high efficiency for detecting low-mass exoplanets and for constraining their masses and orbital parameters, thanks to its sub-1\,m/s RV stability \citep{Pepe2011}.

Special attention is paid to the treatment of {\it Spitzer} systematic noises and to their influence on the results.  Our motivation is also to better understand the limitations of space IR observations for the study of exoplanet atmospheres. Such an effort is especially important  in the context of the future launch of JWST \citep{Seager2009}, in order to optimise the use of its capacities for the studies of other worlds.

In Sect. 2 we present all the observations obtained in the {\it Spitzer} programs targeting GJ\,436b and the way we performed their data reduction. The new HARPS RVs are presented in the same section. Section 3 summarises the  analysis of the {\it Spitzer} data and the RV measurements. Section 4 discusses the possible evidence of companions, while Sect. 5 presents an atmospheric model based on our emission and transmission spectra. We discuss our results in Sect. 6, before concluding in Sect. 7.

\begin{table*}[t]
\scriptsize
\centering
\caption{\label{tIRAC} Presentation of all the {\it Spitzer} data of GJ\,436 used in this paper. They are classified in ascending order of the bandpass wavelength and observing date. The first column gives the bandpass transmission centre of the instrument/channel, the second the eclipse nature of the event, the third the program ID and its corresponding Astronomical Observation Request (AOR) number in column 4. The content of the following columns results from our analysis and is mainly linked to the data analysed with aperture photometry. Column 5 provides the chosen aperture photometry radii. In column 6 the ``background contribution'' indicates the relative sky background contribution in the chosen aperture photometry. Columns 7 and 8 give the horizontal and vertical average coordinates of the point-spread function (PSF) centre in fractional pixel units, considering the bottom left corner to have (0,0) for coordinates. The centre is computed by fitting a 2D elliptical Gaussian for the bandpass ranging from 3.6 to 8 $\mu$m. Otherwise, it is done during the deconvolution process. Note that we do not give the PSF centre at 24 $\mu$m since there are 14 different locations and that they do not depend on the AOR. Each of the following columns respectively gives the baseline function selected for our global modelling (see Sect.~3.3), the $\beta_{w}$ and $\beta_{r}$ errors rescaling factors, the time interval used to compute $\beta_{r}$, and the average total flux of the system. For the baseline function, $p(\epsilon^{N})$ respectively denotes a $N$-order polynomial function of the time ($\epsilon = t$), of the logarithm of time ($\epsilon = l$), and of the PSF $x-$ and $y-$ positions ($\epsilon=$[$xy$]) and widths ($\epsilon=w_{x}$ \& $\epsilon=w_{y}$). }
\begin{tabular}{ccccccccccccc}
\hline\hline
Bandpass& Eclipse & Program & AOR & Aperture  & Background  & \multicolumn{2}{c}{PSF centre} & Baseline & $\beta_{w}$  & $\beta_{r}$ & $T_{\beta_{r}}$ &Average  \\ \noalign{\smallskip}
 &  nature & ID &  & radius & contribution  & x & y & model  &  & &  & total flux \\ \noalign{\smallskip}
($\mu$m) &  & & & (pixels) &   (\%) & (pixels) & (pixels) &  &  &  &(min) & (mJy) \\ \noalign{\smallskip}
\hline  \noalign{\smallskip}
3.6 & Occultation 	& 40685 & 24882688 & 1.9 & 0.04 & 15.67 & 15.97 & $p($[$xy$]$^{3}+w_{x}^{1}+w_{y}^{1}+t^{2})$ &0.95 & 1.24 & 15 & 1269.7 $\pm$ 2.3 \\ \noalign{\smallskip} %
3.6 & Transit 		& 50051 & 28894208 & 2.4 & 0.05 & 16.41 & 15.85 & $p($[$xy$]$^{1})$ &1.00 & 1.60 & 60 & 1254.0 $\pm$ 1.7 \\ \noalign{\smallskip} %
3.6 & Transit 		& 50051 & 28894464 & 3.0 & 0.12 & 16.11 & 16.11 & $p($[$xy$]$^{2}+w_{x}^{1}+w_{y}^{1}+l^{1}+t^{2})$ &0.97 & 1.00 &5& 1269.9 $\pm$ 2.2 \\ \noalign{\smallskip} %
3.6 & N/A			& 60003 & 38807296 & 2.5 & 0.29 & 16.84 & 26.49 & $p($[$xy$]$^{2}+w_{x}^{2}+w_{y}^{2}+l^{2}+t^{4})$ &1.25 & 2.87 &40& 1253.8 $\pm$ 9.9 \\ \noalign{\smallskip} %
3.6 & Occultation	& 60003 & 40848384 & 2.2 & 0.01 & 16.27 & 15.77 & $p($[$xy$]$^{3}+w_{x}^{3}+w_{y}^{3}+l^{1}+t^{4})$ &1.12 & 2.88    &35& 1274.7 $\pm$ 8.2 \\ \noalign{\smallskip} %
4.5 & Occultation 	& 40685 & 24882944 & 1.9 & 0.07 & 15.74 & 15.95 & $p($[$xy$]$^{1}+w_{x}^{1}+t^{1})$ &1.04 & 1.84 &65& 841.10 $\pm$  3.02 \\ \noalign{\smallskip} %
4.5 & Transit 		& 50051 & 28894720 & 2.4 & 0.12 & 15.99 & 15.92 & $p($[$xy$]$^{1}+w_{x}^{1}+w_{y}^{2}+l^{1})$ &1.10 & 1.36 &40& 848.50 $\pm$  3.07 \\ \noalign{\smallskip} %
4.5 & Transit 		& 50051 & 28894976 & 2.2 & 0.10 & 15.55 & 15.98 & $p($[$xy$]$^{2}+w_{y}^{1}+l^{1}+t^{1})$ &1.06 & 1.17 &25& 841.91 $\pm$  3.61 \\ \noalign{\smallskip} %
4.5 & N/A			& 541     & 38702592 & 2.2 & 0.13 & 15.70 & 15.91 & $p($[$xy$]$^{1}+w_{x}^{2}+w_{y}^{3}+l^{1})$ &1.09 & 1.20 &120& 857.20 $\pm$  2.20 \\ \noalign{\smallskip} %
4.5 & N/A			& 60003 & 38808064 & 3.5 & 0.21 & 15.92 & 26.36 & $p($[$xy$]$^{1}+w_{x}^{1}+w_{y}^{1}+t^{1})$ &1.04 & 1.42 &40& 854.46 $\pm$  2.20  \\ \noalign{\smallskip} %
4.5 & Occultation 	& 60003 & 40848128 & 2.2 & 0.13 & 15.68 & 16.29 & $p($[$xy$]$^{1}+w_{x}^{2}+w_{y}^{2}+l^{1}+t^{1})$ &1.12 & 1.00 &5& 847.55 $\pm$  1.90 \\ \noalign{\smallskip} %
4.5 & N/A         	 	& 70084 & 42614016 & 2.4 & 0.14 & 15.70 & 16.01 & $p($[$xy$]$^{1}+w_{x}^{1}+w_{y}^{1})$ & 1.21 & 1.06 &30&  856.15 $\pm$  1.45 \\ \noalign{\smallskip} %
5.8 & Occultation 	& 40685 & 24883200 & 2.8 & 0.28 & 15.85 & 15.39 & $p($[$xy$]$^{2}+t^{1})$ &1.26 & 1.00 &5& 374.57 $\pm$ 0.84 \\ \noalign{\smallskip} %
8 & Transit 		      & 30129 & 23515648 & 3.2 & 2.90 & 15.78 & 15.96 & $p($[$xy$]$^{1}+l^{2})$ &1.14 & 1.00 &5& 207.66 $\pm$  1.29 \\ \noalign{\smallskip} %
8 & Occultation   	& 30129 & 23618304 & 2.4 & 2.00 & 15.74 & 15.62 &$p($[$xy$]$^{1}+l^{1}+t^{2})$ & 1.15 & 1.14 &60& 205.31 $\pm$  0.95 \\ \noalign{\smallskip} %
8 & Occultation		& 50734 & 26812928 & 2.6 & 0.98 & 16.05 & 15.66 &$p($[$xy$]$^{1}+w_{y}^{1}+l^{1})$ & 1.03 & 1.00&5 & 205.44 $\pm$ 1.13 \\ \noalign{\smallskip} %
8 & Occultation		& 50734 & 27604736 & 3.4 & 1.60 & 15.62 & 15.36 & $p($[$xy$]$^{2}+l^{1})$ &1.18 & 1.33 &45& 205.81 $\pm$  1.03 \\ \noalign{\smallskip} %
8 & Occultation		& 50734 & 27604992 & 3.5 & 1.92 & 16.11 & 15.53 &$p(l^{1})$ & 1.17 & 1.37 &15& 205.58 $\pm$ 1.07 \\ \noalign{\smallskip} %
8 & Occultation		& 50734 & 27605248 & 3.4 & 1.88 & 15.92 & 15.73 &$p($[$xy$]$^{1}+l^{1})$ & 1.14 & 1.15 &20& 205.48 $\pm$  1.17 \\ \noalign{\smallskip} %
8 & Occultation		& 50056 & 27863296 & 4.1 & 5.50 & 15.55 & 15.75 & $p($[$xy$]$^{1}+l^{1}+t^{2})$ &1.30 & 1.07 &20& 205.90 $\pm$  1.03 \\ \noalign{\smallskip} %
8 & Transit		      & 50056 & 27863552 & 4.1 & 5.70 & 15.53 & 15.52 &$p($[$xy$]$^{1}+w_{x}^{1}+w_{y}^{1})$ & 1.38 & 1.13 &5& 205.94 $\pm$ 0.94\\ \noalign{\smallskip} %
8 & Occultation		& 50056 & 27863808 & 4.3 & 6.41 & 15.53 & 15.27 &$p($[$xy$]$^{1}+t^{2})$ & 1.30 & 1.02 &5& 205.99 $\pm$ 0.98 \\ \noalign{\smallskip} %
8 & N/A			      & 50056 & 27864064 & 2.8 & 3.24 & 15.52 & 15.24 & $p($[$xy$]$^{2})$ &1.15 & 1.00 &5&  205.97 $\pm$ 1.08 \\ \noalign{\smallskip} %
8 & Transit  		& 50051 & 28895232 & 2.7 & 2.06 & 15.85 & 15.55 & $p($[$xy$]$^{2}+l^{2}+t^{1})$ &1.16 & 1.00 &10& 205.36 $\pm$  1.10 \\ \noalign{\smallskip} %
8 & Occultation		& 50734 & 28969472 & 3.5 & 2.87 & 16.18 & 16.37 &$p(l^{1})$ & 1.23 & 1.02 &5& 205.77  $\pm$ 1.20 \\ \noalign{\smallskip} %
8 & Occultation		& 50734 & 28969728 & 3.3 & 2.38 & 15.75 & 16.01 & $p($[$xy$]$^{1}+w_{y}^{1}+l^{1}+t^{1})$ &1.29 & 1.00 &5& 205.71 $\pm$ 1.30 \\ \noalign{\smallskip} %
8 & Occultation		& 50734 & 28969984 & 2.5 & 1.56 & 15.93 & 15.69 & $p($[$xy$]$^{1}+l^{1}+t^{1})$ &1.15 & 1.00 &5&  205.73 $\pm$ 1.22 \\ \noalign{\smallskip} %
8 & Transit 		      & 50051 & 28895488 & 3.0 & 2.00 & 15.52 & 15.80 & $p($[$xy$]$^{2}+l^{1}+t^{1})$ &1.23 & 1.00 &5&  205.15 $\pm$  0.99 \\ \noalign{\smallskip} %
8 & Occultation		& 50734 & 28970240 & 3.5 & 2.40 & 16.12 & 15.62 & $p($[$xy$]$^{1}+l^{1})$ &1.22 & 1.00 &5&  205.60 $\pm$ 1.14 \\ \noalign{\smallskip} %
16 & Occultation 	& 40685 & 23799552 & - & -    & 21.99 & 27.89 & $p($[$xy$]$^{2}+l^{2}+t^{3})$ &1.86 & 1.00 &5& 85.99 $\pm$ 0.31 \\ \noalign{\smallskip} %
24 & Occultation 	& 40685 & 23799296 & - & - & -  & - & - &3.09 & 1.13 &5& 37.33 $\pm$ 0.40 \\ \noalign{\smallskip}
     &  			&            & 23798784 & - & - & - & - & -  & -\\ \noalign{\smallskip}
     &  			&            & 23800320 & - & - & - & - & - & - \\ \noalign{\smallskip}
     &  			&            & 23801856 & - & - & - & - & - & - \\ \noalign{\smallskip}
     &  			&            & 23801600 & - & - & - & - & - & - \\ \noalign{\smallskip}
     &  			&            & 23801344 & - & - & - & - & - & - \\ \noalign{\smallskip}  

\hline  \noalign{\smallskip}
\end{tabular}
\end{table*}

\onllongtab{2}{
\begin{longtable}[c]{ccc}
\caption{\label{HARPS} Radial velocities of GJ\,436.} \\ \noalign{\smallskip} 
\hline\hline \noalign{\smallskip} 
BJD$_{\rm TT}$ - 2\,450\,000.0 & RV (km s$^{-1}$) & $\sigma_{\rm RV}$ (km s$^{-1}$) \\  \noalign{\smallskip} 
\hline
\endfirsthead
\caption{continued.} \\ \noalign{\smallskip} 
\hline\hline \noalign{\smallskip} 
BJD$_{\rm TT}$ - 2\,450\,000.0 & RV (km s$^{-1}$) & $\sigma_{\rm RV}$ (km s$^{-1}$) \\  \noalign{\smallskip} 
\hline
\endhead
\hline 
\endfoot
3760.83575 &  9.774490 &  0.001110 \\  
3761.84061 &  9.807520 &  0.001110 \\  
3762.82805 &  9.782330 &  0.000900 \\  
3763.84203 &  9.780590 &  0.000890 \\  
3765.80411 &  9.775230 &  0.000940 \\  
3785.76121 &  9.809860 &  0.000930 \\  
3788.77940 &  9.800960 &  0.000900 \\  
4122.84306 &  9.776460 &  0.000910 \\  
4135.83242 &  9.781010 &  0.000920 \\  
4140.82196 &  9.788840 &  0.001030 \\  
4142.83025 &  9.808690 &  0.000910 \\  
4166.75093 &  9.806690 &  0.001140 \\  
4172.74297 &  9.782380 &  0.001000 \\  
4194.70900 &  9.777700 &  0.001230 \\  
4197.67745 &  9.791090 &  0.001090 \\  
4199.67093 &  9.774910 &  0.000970 \\  
4202.66419 &  9.777130 &  0.000950 \\  
4228.58285 &  9.778490 &  0.001040 \\  
4230.48575 &  9.797180 &  0.001170 \\  
4230.49514 &  9.796850 &  0.001280 \\  
4230.50134 &  9.794220 &  0.001800 \\  
4230.50510 &  9.793220 &  0.001760 \\  
4230.50900 &  9.796050 &  0.001610 \\  
4230.51279 &  9.795530 &  0.001550 \\  
4230.51650 &  9.798750 &  0.001690 \\  
4230.52037 &  9.796760 &  0.001680 \\  
4230.52431 &  9.794440 &  0.002520 \\  
4230.52814 &  9.795850 &  0.001970 \\  
4230.53201 &  9.796930 &  0.001700 \\  
4230.53564 &  9.794730 &  0.001540 \\  
4230.53947 &  9.791850 &  0.001510 \\  
4230.54345 &  9.793580 &  0.001500 \\  
4230.54718 &  9.792180 &  0.001510 \\  
4230.55108 &  9.793940 &  0.001490 \\  
4230.55499 &  9.792370 &  0.001530 \\  
4230.55882 &  9.793830 &  0.001410 \\  
4230.56257 &  9.793770 &  0.001370 \\  
4230.56648 &  9.793990 &  0.001440 \\  
4230.57016 &  9.795420 &  0.001450 \\  
4230.57410 &  9.793690 &  0.001290 \\  
4230.57783 &  9.795000 &  0.001290 \\  
4230.58159 &  9.794400 &  0.001470 \\  
4230.58552 &  9.793470 &  0.001320 \\  
4230.58943 &  9.794290 &  0.001310 \\  
4230.59315 &  9.793000 &  0.001440 \\  
4230.59713 &  9.792140 &  0.001390 \\  
4230.60078 &  9.790910 &  0.001620 \\  
4230.60464 &  9.793860 &  0.001570 \\  
4230.60854 &  9.793070 &  0.001760 \\  
4230.61237 &  9.792890 &  0.001780 \\  
4230.61624 &  9.792660 &  0.001990 \\  
4230.61995 &  9.792060 &  0.001930 \\  
4230.62385 &  9.797390 &  0.001890 \\  
4230.62773 &  9.790550 &  0.001540 \\  
4230.63151 &  9.791650 &  0.001500 \\  
4230.63534 &  9.794210 &  0.001380 \\  
4230.63951 &  9.792870 &  0.001370 \\  
4230.64327 &  9.791080 &  0.001340 \\  
4230.64710 &  9.792050 &  0.001360 \\  
4230.65100 &  9.791110 &  0.001320 \\  
4230.65472 &  9.791060 &  0.001360 \\  
4230.65863 &  9.790150 &  0.001400 \\  
4234.55338 &  9.787160 &  0.001180 \\  
4253.54525 &  9.805840 &  0.001060 \\  
4254.51133 &  9.785900 &  0.001180 \\  
4255.51652 &  9.776930 &  0.001160 \\  
4259.48729 &  9.794650 &  0.000860 \\  
4291.48864 &  9.786670 &  0.001770 \\  
4292.47496 &  9.776630 &  0.000970 \\  
4293.45165 &  9.807350 &  0.000840 \\  
4294.45008 &  9.780610 &  0.001020 \\  
4296.47679 &  9.797980 &  0.001300 \\  
4297.45381 &  9.773530 &  0.000970 \\  
4478.85281 &  9.800580 &  0.000950 \\  
4478.87832 &  9.800520 &  0.000940 \\  
4479.86544 &  9.774800 &  0.001060 \\  
4479.87612 &  9.776220 &  0.001020 \\  
4480.85907 &  9.802150 &  0.000970 \\  
4480.87048 &  9.803650 &  0.001080 \\  
4481.86319 &  9.787750 &  0.000980 \\  
4481.87376 &  9.787880 &  0.000940 \\  
4482.85401 &  9.779690 &  0.001060 \\  
4482.86479 &  9.777000 &  0.001040 \\  
4483.85232 &  9.810150 &  0.001000 \\  
4483.86342 &  9.810410 &  0.000940 \\  
4485.85751 &  9.791110 &  0.001000 \\  
4485.86862 &  9.791410 &  0.001000 \\  
4486.85432 &  9.797240 &  0.001050 \\  
4486.86446 &  9.797920 &  0.001130 \\  
4487.85213 &  9.776690 &  0.001200 \\  
4487.86293 &  9.776470 &  0.001060 \\  
4488.85500 &  9.808410 &  0.001100 \\  
4522.83874 &  9.790160 &  0.001230 \\  
4523.78988 &  9.801300 &  0.001070 \\  
4524.79013 &  9.774390 &  0.000900 \\  
4525.81324 &  9.805200 &  0.001110 \\  
4526.76364 &  9.789100 &  0.000900 \\  
4527.75196 &  9.775780 &  0.000960 \\  
4528.71729 &  9.809030 &  0.000930 \\  
4529.81240 &  9.781870 &  0.000990 \\  
4530.77746 &  9.788820 &  0.000970 \\  
4548.70327 &  9.775930 &  0.000880 \\  
4549.69758 &  9.807770 &  0.001150 \\  
4551.71918 &  9.780560 &  0.000950 \\  
4553.69816 &  9.777170 &  0.000800 \\  
4557.65838 &  9.809100 &  0.001110 \\  
4562.64495 &  9.797950 &  0.000990 \\  
4564.64742 &  9.777990 &  0.001220 \\  
4567.56615 &  9.779660 &  0.000820 \\  
4567.73507 &  9.787220 &  0.000860 \\  
4568.57079 &  9.807880 &  0.000890 \\  
4568.69152 &  9.805640 &  0.000940 \\  
4569.62883 &  9.775720 &  0.000860 \\  
4570.63780 &  9.799210 &  0.000920 \\  
4571.60962 &  9.792680 &  0.001020 \\  
4593.55210 &  9.774620 &  0.000960 \\  
4610.47902 &  9.809060 &  0.001120 \\  
4610.56287 &  9.812700 &  0.001080 \\  
4611.55737 &  9.784050 &  0.001030 \\  
4616.47724 &  9.796650 &  0.001410 \\  
4832.87687 &  9.807660 &  0.001190 \\  
4848.87131 &  9.805790 &  0.001020 \\  
4854.84706 &  9.782360 &  0.001050 \\  
4878.78979 &  9.778840 &  0.000940 \\  
4879.79209 &  9.790510 &  0.001380 \\  
4880.79826 &  9.795860 &  0.001690 \\  
4880.81312 &  9.797910 &  0.001240 \\  
4881.78795 &  9.775030 &  0.000990 \\  
4882.78434 &  9.806450 &  0.001100 \\  
4883.79460 &  9.787100 &  0.000920 \\  
4884.77339 &  9.777030 &  0.001400 \\  
4885.79300 &  9.807700 &  0.000920 \\  
4886.77008 &  9.777430 &  0.001080 \\  
4913.68867 &  9.777280 &  0.000940 \\  
4914.70871 &  9.813120 &  0.001000 \\  
4915.68788 &  9.783630 &  0.001060 \\  
4916.69305 &  9.786200 &  0.000950 \\  
4917.68034 &  9.805840 &  0.001020 \\  
4918.72808 &  9.777910 &  0.000990 \\  
4919.66311 &  9.801600 &  0.001030 \\  
4920.70757 &  9.793540 &  0.000960 \\  
4932.64335 &  9.788420 &  0.001300 \\  
4933.63867 &  9.799470 &  0.001070 \\  
4934.62726 &  9.776640 &  0.001010 \\  
4936.63242 &  9.791100 &  0.001500 \\  
4937.61224 &  9.780310 &  0.001200 \\  
4938.62352 &  9.810220 &  0.000960 \\  
4939.64008 &  9.781560 &  0.001430 \\  
4940.61510 &  9.793920 &  0.001370 \\  
4941.62296 &  9.799100 &  0.000980 \\  
4946.60518 &  9.807600 &  0.000890 \\  
4949.58281 &  9.797480 &  0.000980 \\  
4950.58399 &  9.775750 &  0.000940 \\  
4953.58922 &  9.784570 &  0.001040 \\  
4954.58379 &  9.808130 &  0.000960 \\  
4955.58240 &  9.779020 &  0.000990 \\  
4956.58363 &  9.801020 &  0.001210 \\  
4990.49882 &  9.778470 &  0.001120 \\  
4991.51076 &  9.810730 &  0.001030 \\  
4993.46558 &  9.788340 &  0.000980 \\  
4998.46541 &  9.779650 &  0.001320 \\  
4999.45630 &  9.807500 &  0.001230 \\  
5272.70295 &  9.780040 &  0.001110 \\  
5275.70795 &  9.773080 &  0.001320 \\  
5277.71133 &  9.789610 &  0.001260 \\  
5278.70343 &  9.779770 &  0.001690 \\  
5279.68465 &  9.807250 &  0.001170 \\  
5280.70084 &  9.777020 &  0.001190 \\  
5283.67801 &  9.775340 &  0.000980 \\  
5287.65616 &  9.811220 &  0.001080 \\  
5292.65024 &  9.812290 &  0.002210 \\  
\end{longtable}
}

\section{Observations and data reduction}
\label{sec:Obs}

{\it Spitzer} made the first transit and occultation observations of GJ\,436b  in June 2007 at 8 $\mu$m under the Target of Opportunity (ToO) program (ID : 30129) proposed by J.\,Harrington. After the publication of several analyses of these data \citep{Gillon2007a,Deming2007,Demory2007}, the occultations of the planet were observed in the other {\it Spitzer} bandpasses in the framework of  ToO program  40685 (PI J.\,Harrington). The resulting emission spectrum was presented in \citetalias{Stevenson2010}.  Later, the General Observer (GO) program of H. Knutson was motivated to detecting  GJ\,436b's phase variation  (ID : 50056). Afterwards, two sets of four consecutive occultations at 8 $\mu$m were planned within G. Laughlin's GO program (ID : 50734,  \citetalias{Knutson2011}) in order to analyse the tidal heating of the planet. Later, H. Knutson searched for transit depth variations in  the 3.6, 4.5, and 8.0~$\mu$m IRAC bands   (ID : 50051, \citetalias{Beaulieu2011}, \citetalias{Knutson2011}). Furthermore, the program 60003 attempted to observe occultations of GJ\,436b at 3.6 and 4.5~$\mu$m. Wrong ephemerides were used, and the occultation phase was not observed. Lately, S. Ballard (ID : 541, \citealt{Ballard2010b}) attempted to identify a third body in the system during an 18-hour observation at 4.5~$\mu$m. Finally, J.\,Harrington et al. obtained two more observations at 3.6 and 4.5~$\mu$m with  Warm {\it Spitzer} during two occultations of GJ\,436b to try to confirm the conclusions given by \citetalias{Stevenson2010} and  \citetalias{Stevenson2012}. 
Table \ref{tIRAC} summarises this extensive {\it Spitzer} dataset according to their scheduling units (called AOR = Astronomical Observation Request in {\it Spitzer} terminology).

In this section we present the photometric data with the different tested data reduction strategies on each instrument and our updated HARPS RVs dataset. For all {\it Spitzer} instruments, we use the images calibrated by the standard {\it Spitzer} pipeline and delivered to the community as Basic Calibrated Data (BCD). Their units are converted from MJy/sr to electrons. We apply the routine\footnote{http://astroutils.astronomy.ohio-state.edu/time/hjd2bjd.html} described by \cite{Eastman2010} to convert Julian days into BJD$_{TT}$ and we transform the IRAC data timestamps following  \citetalias{Knutson2011} from BJD$_{UTC}$ to BJD$_{TT}$ at mid-exposure time. 

\subsection{{\it MIPS} observations}

On 2008 January 4, the 24 $\mu$m channel of MIPS was used for 6 hours to observe GJ\,436. During every cycle, MIPS acquired five times the images of the source at 14 different locations of the detector distributed on 2 columns. At the beginning of every cycle one extra exposure of one second shorter than the specified exposure time\footnote{see \S 3.1.1 of the {\it{MIPS instrument handbook}}} was done for each column to obtain calibration data.  

We discard the two extra exposures  of every cycle  and only analyse the science data, i.e. the 1680 9.96-second exposures  from the 1728 available BCD images (version S17.0.4). The dithering scheme allows us to remove the sky background. For a given frame, all frames having a maximum time difference of X min and a different dither position are combined to produce a master sky frame. We test X = 3.6, 7.2, 10.8, and 14.4~minutes. We finally select the value X = 14.4~minutes as it leads to the best photometric quality. We also test photometric reductions performed without preliminary background subtraction, but they lead to much poorer photometry  because of significantly larger correlated noise and scatter.

For the flux measurements, we test aperture photometry but the results are not satisfactory. \cite{Deming2005}, \cite{Knutson2009mips} and \citetalias{Stevenson2010} previously demonstrated that methods using the Point Spread Function (hereafter PSF) should be preferred to aperture photometry when dealing with MIPS data. In this context, we choose to apply the partial deconvolution photometry method DECPHOT described by \citet{decphot1,decphot2} and based on the MCS image deconvolution method \citep{Magain1,Magain2}. The word ``partial'' refers to the fact that we use a partial PSF $s({\bf x})$, instead of the total PSF $t({\bf x})$, which can then be represented as the convolution of the resulting PSF in the deconvolved image $r({\bf x})$ by the partial PSF~:

\begin{equation}
t({\bf x}) = s({\bf x}) \ast r({\bf x})
\end{equation} 
where $\ast$ stands for the convolution operator and ${\bf x}$ for the pixel position in the array. $r({\bf x})$ is a Gaussian function chosen in such a way that the final result  is well sampled.

DECPHOT relies on the minimisation of the following merit function:

\begin{equation}
\label{eq:meritfct}
S = \displaystyle\sum_{i=1}^{N}\frac{1}{\sigma_{i}^{2}} (d_{i}-b_{i}- [s ({\bf x}) \ast f({\bf x})]_{i})^{2} + \lambda H(s)
\end{equation}  
where $\sigma_{i}$ is the standard deviation at pixel $i$. The values of $d_{i}$, $b_{i}$ and $f_{i}$ respectively stand for the observed light, the sky and the deconvolved light level at pixel $i$. The residual image background is thus simultaneously  retrieved during the deconvolution process. It is represented by a 2-dimensional second order polynomial (6 free parameters). $H(s)$ is a smoothing constraint on the PSF that is introduced to regularise the solution and $\lambda$ is a Lagrange parameter.
If the image is composed of point sources, the deconvolved light distribution $f$ may be written:
\begin{equation}
f({\bf x}) = \sum_{j=1}^{M}a_{j}r({\bf x}-{\bf c}_{j})
\end{equation} 
where $a_{j}$ and ${\bf c}_{j}$ are free parameters corresponding to the intensity and position of the point source number $j$.
A crucial point in the deconvolution process is the partial PSF construction: the more accurate the partial PSF, the better the deconvolution \citep[e.g.,][]{Letawe2008}.  We derive the partial PSFs  by deconvolving the models made with the {\it Spitzer} Tiny Tim Point Spread Function program \citep{Krist2002} available on the {\it Spitzer} website for the different array locations. We  compute the statistical weight of every pixel for every image to discard cosmic hits and other deviant pixels during the deconvolution process. Finally we deconvolve each image with its corresponding partial PSF, solving in the process for the star's flux. 

\begin{figure*}
\centering
  \includegraphics[width=18cm,trim=0 0 0 20,clip=true]{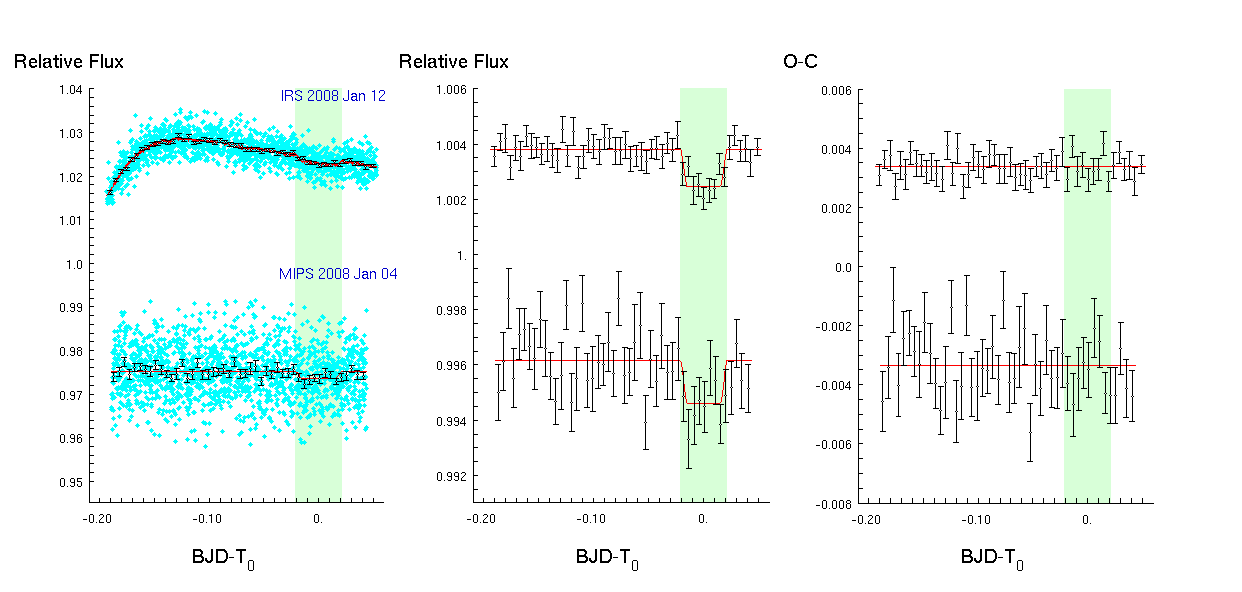}
  \caption{\label{lcoccim} Secondary eclipses of GJ\,436b observed with the IRS (top) and MIPS (bottom) instruments.  Relative flux offsets are applied to datasets for clarity. On the \textbf{left}, raw data are represented by cyan dots for the unbinned data, and by black dots for binned data per interval of 7 min with their error bars. The superimposed red line is the best-fitting model and includes the transit model. On the \textbf{middle}, the same binned data  divided by the best-fitting baseline model reveal the occultation shape. The eclipse model is superposed in red. The \textbf{right} panel displays the binned residuals from the same dataset. The shaded green area of every panel shows the eclipse event.}
\end{figure*}

In the context of this analysis, the main advantage of DECPHOT is to optimally separate the light source from the (residual) complex background. This is crucial for high-precision infrared photometry in the high background regime \citep[see][]{Gillon2009corot1}.   In addition to the preliminary sky subtraction, we have to include an analytical model  for the residual sky background (cf. $b_i$ in Eq.~\ref{eq:meritfct}) in the modelling of the sky-subtracted images during the deconvolution process.  A scalar plane fit is sufficient to model the background close to the stellar source. This highlights low frequency variations of the sky structure on short timescales.   The raw and the fitted light curves are presented in Fig.~\ref{lcoccim} and in the following section. As noticed by \cite{Knutson2009mips}, the flux appears constant at the beginning of the observation and does not  sharply rise as would be the case with the presence of the detector `ramp' effect (see Sect. \ref{subsec:Baseline modelling}). 

\subsection{{\it IRS} observations}

On 2008 January 12, GJ\,436 was monitored by IRS in the 16~$\mu$m peak-up imaging mode over a period of 6 hours with an exposure time of 6.29 seconds. This program was sequenced without any dithering in three parts of which the middle consisted in the observation of GJ\,436 while the two others targeted empty areas on the sky in order to obtain a high-SNR map of the background. 

We discard the sky images because the photometric precision is not improved using them. We perform aperture photometry and partial deconvolution on the 1580 BCD images  (version S17.2) and conclude  as \cite{Knutson2009irs} and \citetalias{Stevenson2010} that the use of the PSF is to be preferred in the data reduction. Only  deconvolution photometry is presented here.

The partial PSF constructed from the over-sampled PSF model given on the {\it Spitzer} website\footnote{http://irsa.ipac.caltech.edu/data/SPITZER/docs/irs/calibrationfiles/ peakuppsfPSF/} does not satisfactorily fit the data. 
We thus modify DECPHOT to simultaneously deconvolve a set of images. This technique attempts to find a unique partial PSF from the images themselves while it allows the position of the point source, its intensity, and  the background to differ from an image to another. We use the same images both to determine the partial PSF and to perform deconvolution photometry. In a first step, we analyse twenty-five random images with our new version of DECPHOT to construct a satisfying partial PSF mode. In a second step, we deconvolve all images, one at a time and using as partial PSF model the result of the first step. The resulting normalised light curve is shown in Fig.~\ref{lcoccim}.

\subsection{IRAC transit and occultation photometry}
\label{subsec:IRACdata}

In the context of the {\it Spitzer} programs 541, 30129, 40685, 50051, 50056, 50734, 60003 and 70084, GJ\,436 was monitored in the four channels of the IRAC camera at 3.6, 4.5, 5.8 and 8 $\mu$m. Those covered 16 occultations and 8 transits of GJ\,436b, including a phase curve. In this work we perform for the first time a uniform analysis of these 29 IRAC time-series.
Our analysis is based on the IRAC BCD images (version S18.18 for the two first channels and S18.25 for the last two). Because GJ\,436 is a bright target for {\it Spitzer} with $K \sim$\,6.1, it was observed in all 4 IRAC channels in subarray mode, meaning that every BCD is a set composed of 64 32$\times$32-pixels subarray images. The telescope was not repointed during the course of the runs to minimise the motion of the star on the array. An exposure time of 0.08\,s was used at 3.6 $\mu$m, 0.08 and 0.32\,s at 4.5 $\mu$m, and 0.32\,s for the two other channels. For more details on these IRAC observations we refer the reader to Table \ref{tIRAC}, \cite{Gillon2007a}, \cite{Deming2007}, \cite{Demory2007}, \cite{Ballard2010b}, \citetalias{Stevenson2010}, \citetalias{Knutson2011}, and \citetalias{Stevenson2012}.

\begin{figure}
\centering
  \includegraphics[width=8.5cm]{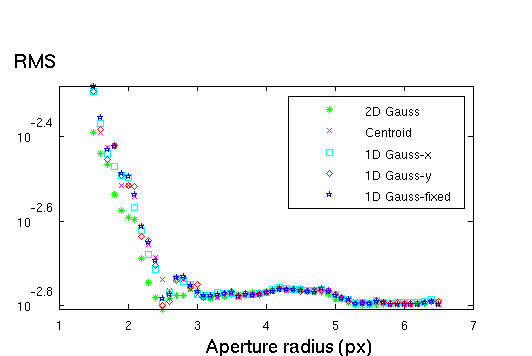}
  \caption{\label{centrage} Evolution of the photometric precision for raw data with the aperture radius according to different centring techniques in a complete dataset (e.g., AOR : 42614016). The 2D Gaussian adjustment (green asterisks) provides the best photometric precision for raw data in comparison to a centroid fit (magenta crosses), a double 1D Gaussian fit from the previously computed {\it x}-or {\it y}-FWHM (cyan boxes and red diamonds respectively), and a 1D Gaussian fit with a fixed FWHM (blue stars), in particular for small aperture radii. }
\end{figure}

We reduce all the IRAC data with our EXOPHOT {\tt PyRAF}\footnote{{\tt PyRAF} is a command language for running IRAF tasks based on the Python scripting language.} pipeline to get raw light curves. We give a general overview of its routines below. For every subarray image, a 2D elliptical Gaussian profile fit is performed to determine the centre of the GJ\,436 PSF. Aperture photometry is then accomplished with the 
{\tt IRAF/DAOPHOT}\footnote{{\tt IRAF} is distributed by the National Optical Astronomy Observatory, which is operated by the Association of Universities for Research in Astronomy, Inc., under cooperative agreement with the National Science Foundation.}  software \citep{Stetson1987}. 
 Other centring approaches are tested (e.g., centroid fit and different double 1D Gaussian adjustments\footnote{A 1-dimensional Gaussian of a given (fixed or pre-computed) FWHM is fitted to the marginal profiles in the $x$- and $y$-directions using non-linear least square techniques. Contrary to a 2D Gaussian fit it adjusts the PSF profile in both directions separately.}) but with lower performance (Fig.~\ref{centrage}).  
The background level is measured in an annulus extending from 12 to 15 pixels from the centre of aperture, and the resulting background flux in the aperture is subtracted from the measured flux for every image.  For every channel, we compute the stellar fluxes in aperture radii ranging from 1.5 to 6 pixels by increments of 0.1. 
We select the aperture radius minimising the scatter of the residuals and their time correlation on the light curves corrected for instrumental and astrophysical effects (see Sect.~\ref{sec:MCMC}). The chosen aperture sizes are all between 1.9 and 4.3 pixels (Table \ref{tIRAC}). Larger apertures lead to larger background contributions while smaller apertures lead to pixelation problems\footnote{Because the fraction of the flux falling into a pixel at the edge of the aperture does not correspond to the fraction of the aperture within the same pixel, the measured flux on each pixel at the aperture border is wrongly evaluated.} and significantly smaller  counts (the IRAC PSF full-width at half maximum ranges from 1.2 to 2 pixels, depending on the channel).

\begin{figure}
\centering
  \includegraphics[width=8.8cm,height=10cm]{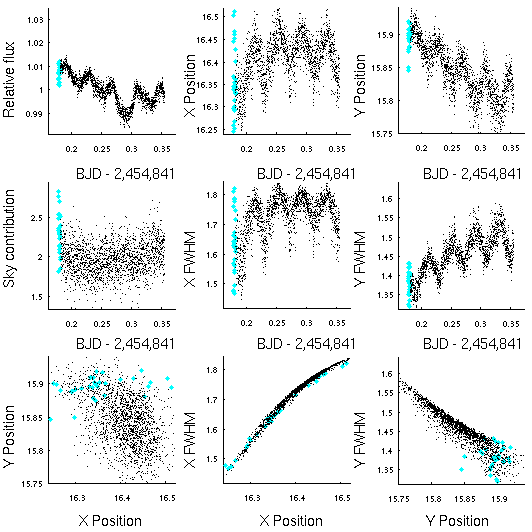}
  \caption{\label{figfwhm}  For illustration, evolution of the parameters measured by our reduction pipeline for a randomly chosen IRAC dataset (at 3.6 $\mu$m, 2009 January 9, AOR : 28894208). From top to bottom and left to right  we display the evolutions of the normalised stellar fluxes, the PSF $x$- and $y$-centres, the background values, and the PSF FWHMs in the $x$- and $y$-directions. The last three panels display the correlation diagrams for several parameters, revealing a clear dependence on the FWHM with the position for both directions.
The first values (large cyan coloured dots) are rejected, as the measured parameters (e.g., sky contribution) are not yet stabilised.}
\end{figure}

For every block of 64 subarray images, we reject the discrepant values for the measurements of the $x$- and $y$-position, and the stellar and background flux using a 3-$\sigma$ median clipping. We generally discard up to 2 measurements from the 64 subarray images. Then the remaining measurements are averaged. We take the photometric error on the mean of the photometric error for every BCD set. 
At this stage, the first measurements of each light curve are discarded if they correspond to deviant values for all or some of the external parameters (detector or pointing stabilisation). In average $\sim$10 min of data are rejected  for each dataset (Fig.~\ref{figfwhm}). Finally we perform a slipping median filtering for each light curve to discard outlier measurements due to, e.g., cosmic hits. 

\begin{figure}
\centering
  \includegraphics[width=8.8cm,trim=0 0 0 0,clip=true]{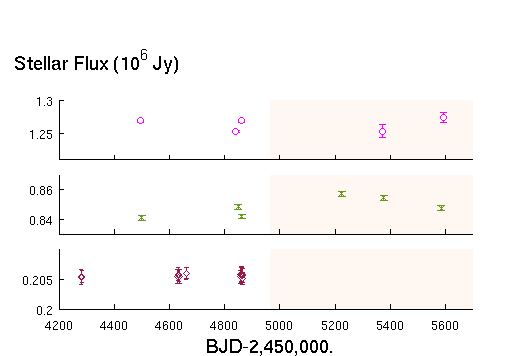}
  \caption{\label{fluxotime}  Stellar flux evolution in the 3.6- (top), 4.5- (middle), and 8.0-$\mu$m (bottom) IRAC channels. The stellar flux slightly fluctuates with time in the 3.6- and 4.5-$\mu$m bandpasses and is constant at 8.0 $\mu$m.   A different inter-pixel map is used for the  Warm mission data. The  Warm {\it Spitzer} phase is shaded in pale pink for clarity. The flux modulation is probably due to the uncorrected intra-pixel effect and to the calibration uncertainties, which are not displayed here for clarity.}
\end{figure}

We also test a photometric reduction of the IRAC data with DECPHOT, without improving results. Still we use DECPHOT to assess the infrared variability of GJ\,436. Indeed, aperture photometry performed on deconvolved images reconvolved by the best-fitting partial PSF model allows us to derive the aperture corrections required for deriving the  observed flux of the star $F_\ast$. For every dataset, we perform this procedure for a large range of apertures. Then we average all the measurements and take the resulting value as the observed flux measurement for the dataset. The error on the mean is considered as its error bar. We apply the colour and inter-pixel corrections\footnote{see \S 4.4 and 4.5 of the {\it{Spitzer Observer's Manual}} and http://irsa.ipac.caltech.edu/data/SPITZER/docs/irac/warmfeatures/}. We do not correct the intra-pixel sensitivity as no complete correction map is available for the  Warm {\it Spitzer} mission at the time of our analysis. The intra-pixel behaviour of the InSb detectors for the Warm {\it Spitzer} mission substantially differs from the one of the cryogenic phase\footnote{see http://irsa.ipac.caltech.edu/data/SPITZER/docs/irac/calibrationfiles/ pixelphasecryo/ and http://irsa.ipac.caltech.edu/data/SPITZER/docs/irac/ calibrationfiles/pixelphase/}, making the correction map for the cryogenic phase unsuitable for all our data. 
The observed stellar flux $F_\ast$ is given in mJy for each IRAC dataset  in Table~\ref{tIRAC}. Their temporal evolution is shown in Fig.~\ref{fluxotime}.  The standard deviations of the fluxes are 0.77, 0.78, and 0.12 \% at 3.6, 4.5, and 8 $\mu$m, respectively. A fraction of the scatter in the shorter wavelengths should come from the absence of intra-pixel correction.  GJ\,436 thus appears to be a stable star in the IR. It is consistent with the nearly constant flux of the star in the optical over a period of five months as observed by \citetalias{Knutson2011} with the Automatic Photoelectric Telescope.  Finally we convert the flux densities  into Vega apparent magnitudes (presented in Sect.~\ref{sec:globanalysis} with the magnitude of the planet) using \cite{Reach2005} magnitude calibrations on the stellar flux measured as it would be falling into a circular aperture radius of 10 pixels.
 
We also test the noise pixel\footnote{http://irsa.ipac.caltech.edu/data/SPITZER/docs/irac/iracinstrument- handbook/5/} method of \cite{Lewis2013} to extract the fluxes of GJ\,436. Based on the use of variable aperture radii, this method aims at minimising the correlation of the fluxes and the PSF centre positions. Our tests for the light curves in the 3.6 and 4.5 $\mu$m shows that this method leads  up to 20\% lower photometric precisions and  larger time correlation of the residual light curves in comparison to traditional fixed aperture photometry and subsequent modelling of the position effects.  It shows similar results only for a few datasets. Consequently, we do not use its resulting light curves in our analysis.

\subsection{HARPS Radial Velocities}
\label{sec:RV}

We used HARPS to record 171 spectra of GJ\,436. Although GJ\,436 was not part of the nominal volume-limited sample of M dwarfs \citep{Bonfils2013} we used the same settings as for the other M stars to record its spectra. We observed without the simultaneous ThAr calibration and relied on the overnight $\lesssim$1~m/s stability of the instrument. This is a better mode for this $V=10.6$ star because it leads to cleaner spectra and still with a photon-noise limited precision of 1~m/s with exposure times of 900 seconds. 

Between 2006 January 25 and 2010 April 6, we obtained 171 measurements of which 44 were taken in a single night and the rest (127) was spread over the whole period. The measurements spread over the years generally have exposure times of 900~s, except some that were made with 1200 and 1800~s to compensate for non-optimal meteorological conditions. They have a median uncertainty of 1.0 m/s. The 44 observations taken on the same night (2007 May 10) aim at measuring the possible Rossiter-McLaughlin effect. They have exposure times of 300 seconds each and radial velocity uncertainties of 1.1$-$1.7 m/s. We note that one of the 127 measurements coincidentally falls during a transit event.  

Extracted and wavelength calibrated spectra are delivered by the standard HARPS pipeline, as well as differential RVs obtained from the cross-correlation between the stellar spectra and a numerical template. To take full benefit of the many spectra of GJ\,436 we merge them to derive a single high signal-to-noise template. We then use this template to re-compute the differential RVs by minimising the $\chi^2$ of the difference between this template and each spectrum. Using a merged stellar spectrum is a common alternative to a numerical template \citep[e.g.,][]{Howarth1997, ZuckerMazeh2006}. It was already used on HARPS by a part of our group (e.g., when we derived the RVs of GJ\,1214, \citealt{Charbonneau2009}) as well as by other groups (e.g., \citealt{Anglada-Escude2012a, Anglada-Escude2012b, Anglada-Escude2013}). Before modelling the data, we subtract its 0.34 m/s/yr secular acceleration \citep{Kurster2003}. Our implementation of the algorithm will be presented in a forthcoming paper (Astudillo et al. in prep.) and the inferred RVs are given in Table~\ref{HARPS}.

\onlfig{5}{
\begin{figure*}
\centering
  \includegraphics[width=18cm,trim=0 15 0 50,clip=true]{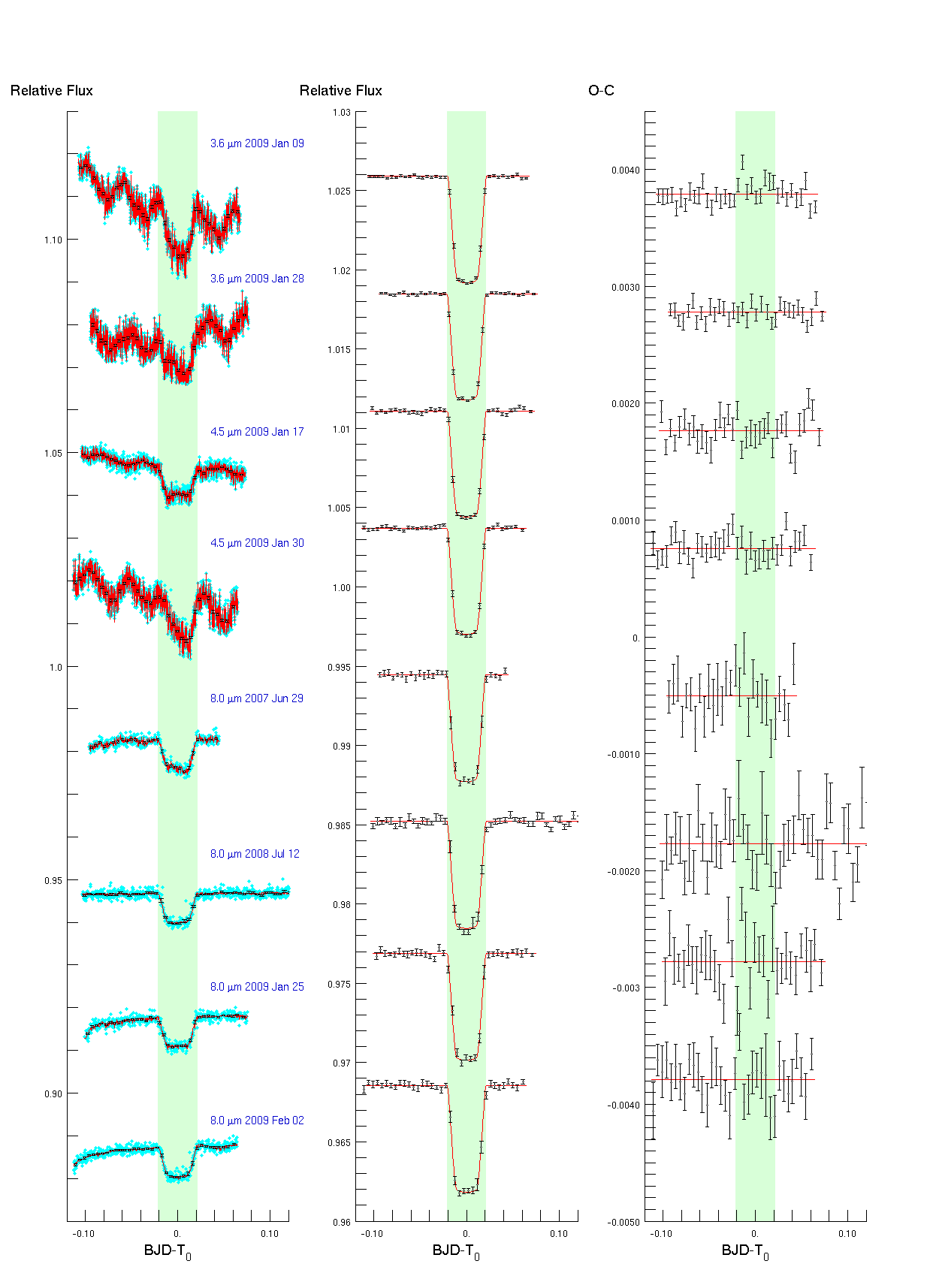}
  \caption{\label{lctra}  Eight transit light curves of GJ\,436b. The two upper panels are the data at 3.6 $\mu$m  from 2009 January 9 and 28. The following two were taken at 4.5 $\mu$m on 2009 January 17 and 30. The lower four were obtained at 8 $\mu$m  on 2007 June 29, 2008 July 12, 2009 January 25 and February 2. The left plot displays the raw binned and unbinned data, the middle one the binned data corrected for instrumental systematics, and the right one the binned residuals. Measurements are binned per interval of 7 min. The light curves are shifted along the $y$-axis for the sake of clarity. The shaded green areas show the transit events.} 
\end{figure*}
}

\onlfig{6}{
\begin{figure*}
\centering
  \includegraphics[width=18cm,trim=0 0 0 50,clip=true]{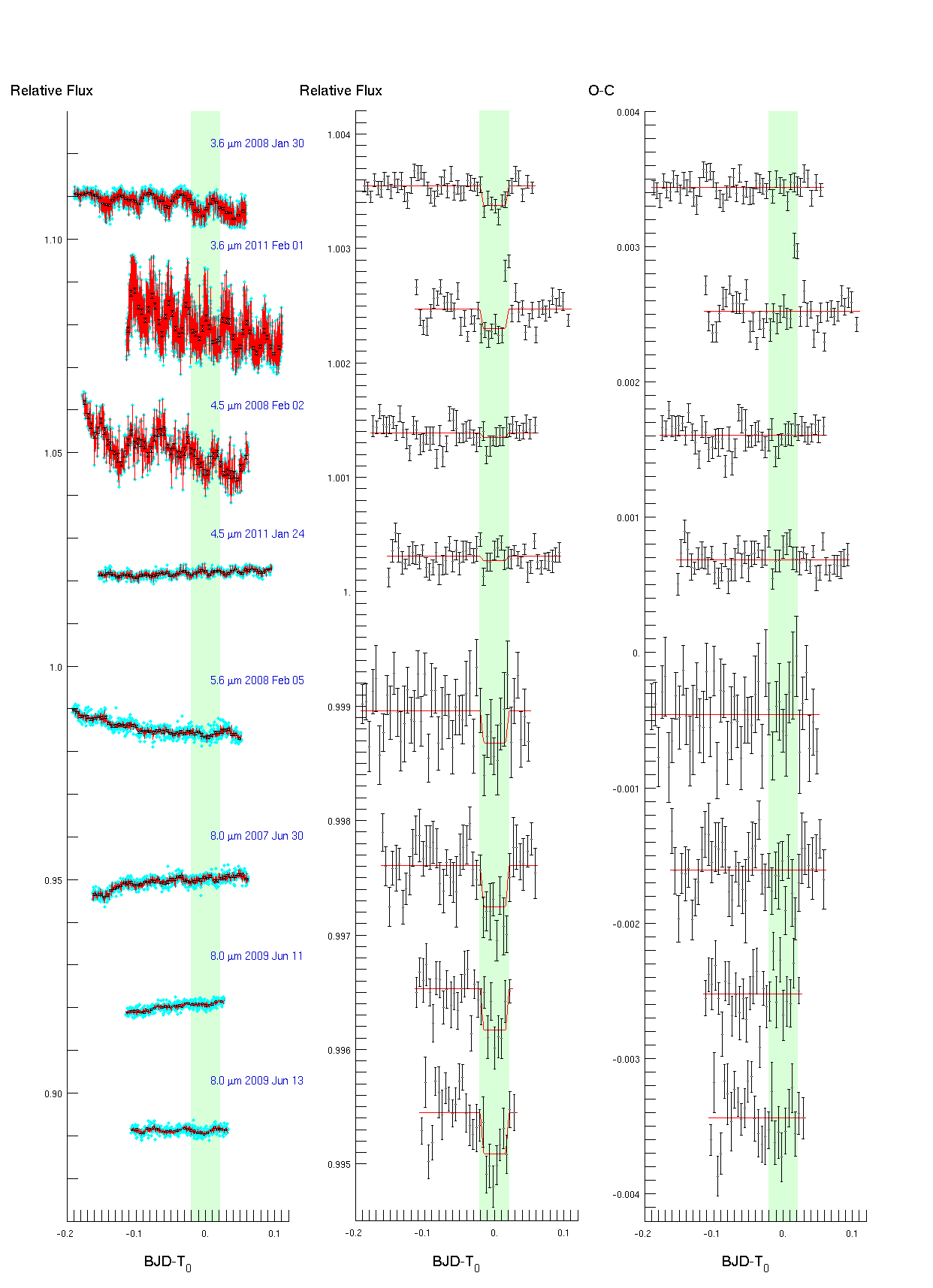}
  \caption{\label{lcocc1}  Eight occultation light curves of GJ\,436b. The two upper plots are the data at 3.6 $\mu$m from 2008 January 30 and 2011 February 1. The following two were obtained with the 4.5 $\mu$m  channel on 2008 February 2 and 2011 January 24. The fifth is the light curve at 5.8 $\mu$m  from 2008 February 5. The last three were taken at 8 $\mu$m  on 2007 June 30, 2008 June 11, and 13. The left plot displays the raw data, the middle one the corrected data, and the right one the residuals. The shaded green areas show the eclipse events.}
\end{figure*}
}

\onlfig{7}{
\begin{figure*}
\centering
  \includegraphics[width=18cm,trim=0 0 0 50,clip=true]{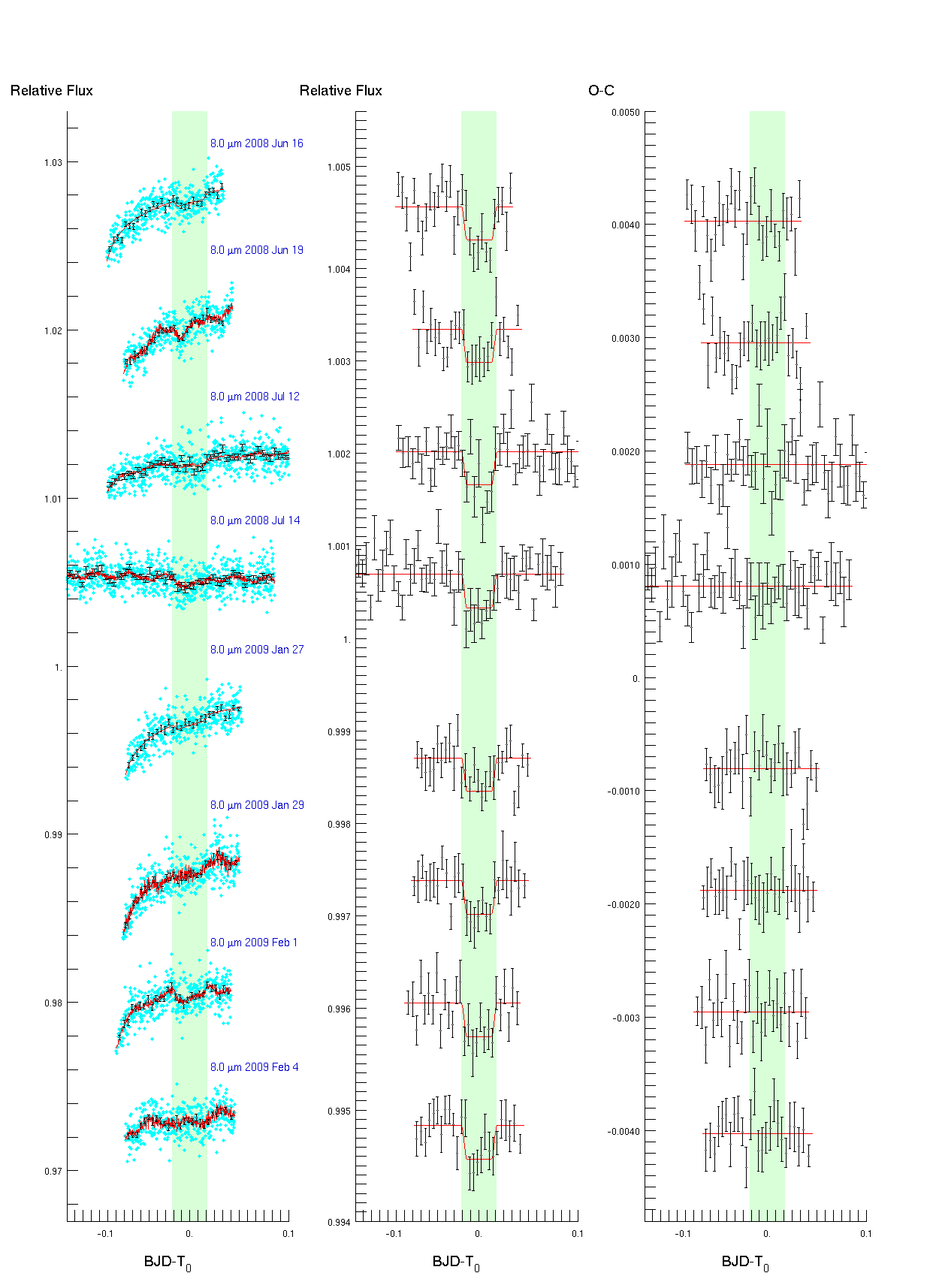}
  \caption{\label{lcocc2}  Same as Fig.~\ref{lcocc1} with eight occultation light curves of GJ\,436b at 8 $\mu$m. They were obtained respectively on 2008 June 16, 19, July 12, 14, 2009 January 27, 29, February 1 and 4.}
\end{figure*}
}

\section{Data Analysis}
\label{sec:MCMC}
We choose to perform a global analysis of our extensive dataset to take full advantage of its observational constraints on the system parameters. Our Bayesian approach of the analysis is based on the use of the Markov Chain Monte Carlo algorithm \citep[MCMC; e.g.,][]{Ford2006}.  

\subsection{Data and model}

In addition to the {\it Spitzer} photometry and the HARPS radial velocity measurements detailed in section~\ref{sec:Obs}, we used the 59 Keck/HIRES RVs presented in \cite{Maness2007} as input data in our global analysis to add further constraints on GJ\,436b's orbital and physical parameters. We also extend this analysis with a total of 29 transit times acquired with the Euler telescope \citep{Gillon2007a}, the Carlos Sanchez telescope \citep{Alonso2008}, the VLT \citep{Caceres2009}, WISE and FWLO \citep{Shporer2009}, telescopes at the Apache Point Observatory \citep{Coughlin2008}, the HST \citep{Pont2009,Bean2008}, and the EPOXI mission \citep{Ballard2010TT}.

An exhaustive description of the MCMC adaptive algorithm applied in this study can be found in \citet{Gillon2010,Gillontrap2012}. We use a model based on a star and a transiting planet on a Keplerian orbit about their centre of mass. For the RVs, a Keplerian 1-planet model is assumed.  A model of the Rossiter-McLaughlin effect is also implemented following \citeauthor{Gimenez2006}' prescription \citeyearpar{Gimenez2006}. We apply the model of \citet{Mandel&Agol2002} to represent the eclipse photometry, assuming a quadratic limb-darkening law for GJ\,436 and  neglecting limb darkening for its planet. For each light curve, an analytical baseline model representing the flux variations of instrumental and stellar origin multiplies the eclipse model (see Sect.~\ref{subsec:Baseline modelling}). 

\subsection{Input stellar parameters}

Thanks to the interferometric measurement from \cite{vonBraun2012}, we have a strong constraint on the stellar radius value of 0.455 $\pm$ 0.018~R$_{\odot}$. Deriving the mass of an M dwarf from a stellar evolution model is not as reliable as for higher mass stars \citep[e.g.,][]{Torres2013}. To best benefit from this high-precision measurement and deduce a posterior distribution for the stellar mass independently from any stellar evolution model, we use the normal distribution $N(0.455,0.018^2)$ as prior distribution for the stellar radius. At each step of the MCMC, the stellar mass  is directly computed from the stellar radius value (constrained by the interferometric prior) and from the stellar density value, which is constrained by the transit light curves \citep{SeagerMallen2003}.

For the atmospheric stellar parameters (effective temperature $T_{\textrm{eff}}$, metallicity [Fe/H], projected rotational velocity $V \sin{i}$, and surface gravity $\log{g}$), normal prior distributions are assumed based on the following arguments: we take a value of 3416~K for the effective temperature \citep{vonBraun2012} derived from spectral energy distribution fitting and adopt  a typical error value of 100~K for the case of low-mass stars \citep[e.g.,][]{Casagrande2008};  
 we use $\log g $ = 4.843 $\pm$ 0.018  from \cite{Torres2007} as a first input before adopting those determined from the stellar mass and radius during the MCMC process; we derive a [Fe/H] of +0.02 inserting  \cite{Schlaufman2010} $\Delta (V-K_s)$ value into \cite{Neves2012} empiric photometry-metallicity relation, which is in good agreement with [Fe/H] = -0.03 from \cite{Bonfils2005} and [Fe/H] = +0.10 from \cite{Schlaufman2010}, and adopt the error of $\pm$0.20 from \citeauthor{Bonfils2005};
finally, we impose $V \sin i < 3$ km.s $^{-1}$ \citep{Delfosse1998} in the MCMC.

For the limb darkening, the two quadratic coefficients $u_1$ and $u_2$ are allowed to float in the MCMC. To minimise the correlations of these model parameters, those coefficients themselves are not used as jump parameters\footnote{Jump parameters are the model parameters that are randomly perturbed at each step of the MCMC.}, only their combinations $c_1 = 2 \times u_1 +u_2$ and $c_2$ = $u_1 - 2 \times u_2$ \citep{Holman2006}. For each transit, $c_1$ and $c_2$ are thus constrained by priors under the control of the normal prior probability distribution functions deduced from \citeauthor{Claret2011}'s tables (\citeyear{Claret2011}).  

\subsection{Baseline model}
\label{subsec:Baseline modelling}
A baseline model aiming at representing systematic effects originating from several sources multiplies each transit/occultation model. It depends on the instrument and wavelength. In addition we include a thermal phase variation model to represent the 8-$\mu$m data. We describe the selection of our baseline model for every AOR in the next subsections.

\subsubsection{The intra-pixel sensitivity and pixelation modelling}
The main source of correlated noise in the IRAC InSb (3.6 and 4.5~$\mu$m) arrays is the dependence on the observed flux on the stellar centroid location on the pixel. These detectors show indeed an inhomogeneous intra-pixel sensitivity, which, combined with the jitter of the telescope and to the poor sampling of the PSF, leads to strong correlation of the measured fluxes with the position of the target’s PSF on the array \citep[e.g.,][and references therein]{Knutson2008}. It is known as the `pixel-phase' effect. 

Our baseline models include two types of low-order polynomials to deal with the pixel-phase effect. The first one uses the prescription of \citet{Desert2009}. It considers the dependence on the fluxes to the $x$- and $y$-positions of the PSF centre with the help of a polynomial that depends on the shift of the centroid centre on the array:
\begin{eqnarray}
F(dx, dy) &= &a_{0} + a_{1}dx + a_{2}dx^{2} + a_{3}dx^{3} + a_{4}dy + a_{5}dy^{2} + a_{6}dy^{3} \nonumber \\
&+& a_{7}dxdy+ a_{8}dxdy^{2}+ a_{9}dx^{2}dy
\end{eqnarray}
where $dx$ and $dy$ are the relative distance of the centroid from the pixel centre and $a_0$ to $a_9$ are free parameters in the fit.

The second one is motivated by the correlation of the FWHM with the PSF centroid location and with the stellar flux variation (Fig.~\ref{figfwhm}).  The following polynomial represents  the dependence on the fluxes to the PSF FWHMs in the {\it x}- and/or {\it y}-directions: 
 
\begin{equation}
\label{eqfwhm}
F(f_x,f_y)  =  a_{10} + a_{11} w_x + a_{12}w_x^{2}+ a_{13}w_x^{3} 
 + a_{14} w_y + a_{15}w_y^{2}+ a_{16}w_y^{3}
\end{equation}
where $w_x$ and $w_y$ respectively stand for the PSF FWHM in the {\it x}- and {\it y}-directions measured by our centring algorithm and $a_{10}$ to $a_{16}$ are free parameters in the fit. 
Modelling this dependence is required for most IRAC InSb light curves.

Only the lowest frequencies of the pixel-phase effect can be modelled with the polynomial approach. For an improved modelling of the effect, 
\cite{Ballard2010b} and \cite{StevensonBLISS} demonstrated the efficiency of building a ``pixel map'' to characterise the intra-pixel variability on a fine grid. Here we  combine  the Bi-Linearly-Interpolated Sub-pixel Sensitivity (BLISS) mapping method presented by \cite{StevensonBLISS} with the position/FWHM polynomial models. The BLISS mapping is performed at every step of the chain after the modelling of the polynomial models. This method maps the intra-pixel sensitivity at high resolution at every step of the MCMC from the data themselves. In our implementation of the method, a sub-pixel-scale grid of $N^2$ knots along the $x$- and $y$-directions slices the sensibility map. We empirically set $N$ in such a way that one average knot is  never associated  to less than 5 measurements to efficiently constrain the sensibility map.  In most cases, we obtain a smaller scatter in the residuals using both parametric models and the sensitivity map.  

\subsubsection{The time varying sensitivity modelling of the detectors}
Another well-documented detector effect, especially important for the Arsenic-doped Silicon (Si:As) material instruments (IRAC 5.8 and 8 $\mu$m, IRS 16 $\mu$m), but also present in the InSb ones, is the increase of the detector response at the start of AORs. The so-called ``ramp''  is attributed to a charge-trapping mechanism resulting in a dependence on the gain of the pixels to their illumination history  \citep[e.g.,][]{Knutson2008}. Following \cite{Charbonneau2008}, we model this ramp with a polynomial of the logarithm of time.

Besides, we also test linear and quadratic functions of time in our baseline models to check time-dependent trends of instrumental and/or stellar origin.


\subsubsection{Phase curve modelling}
\label{sec:phasecurve}

As thermal emission is isotropic, the emission of an atmospheric layer is analogous to the scattering of incident light by a Lambert surface \citep{Seager2010}. Assuming that one specific atmospheric layer dominates the emission at a given wavelength, the thermal phase curve of a pseudo-synchronised planet on an eccentric orbit can be modelled by the following function:

\begin{equation}
\label{eqphascurve}
 F_r = A \left( \frac{r_m}{r} \right)^2 \frac{(\pi - \alpha) \cos(\alpha)+\sin(\alpha)}{\pi}  
\end{equation}
where $A$ is the thermal day-night contrast for a circular orbit, $r$ is the distance between the star and the planet, and $r_m$ is its average. The last fraction models the phase variation of a Lambert sphere  \citep{Russell1916}, with $\alpha$ being the phase angle. It can be easily found with the relation $\cos(\alpha)=\sin(\omega+f) \sin(i)$ \citep[e.g.][]{Sudarsky2005}, where $i$ is the orbital inclination, $\omega$ is the argument at periapsis, and $f$ is the true anomaly.
 This baseline model alone supposes a constant stellar flux during the observation time but takes into account the received stellar radiation variation with time according to the planet-star distance \citep[e.g.][]{KaneGelino2010,Cowan2011}. The phase curve modelling implies the insertion of $A$ and the phase offset in the phase angle as jump parameters.

\subsection{Fitting procedure}
A preliminary evaluation of all the necessary input parameters in our MCMC code is done before performing a global analysis of the GJ\,436 dataset. Furthermore, the multiple observations within the same filters allow us to search for variability in the eclipse parameters. 

\subsubsection{Model selection and error rescaling}

For each light curve (corresponding to a specific AOR), we test a large range of baseline models and look for the minimisation of the Bayesian Information Criterion \citep[BIC; e.g.,][]{Schwarz1978} to select the simplest model that best fits the data (Occam's razor). We aim at detecting the thermal phase curve at 8 $\mu$m prior to the global analysis. However its amplitude is not significant at a 3-$\sigma$ level and the baseline model for the phase curve modelling (Eq.~ \ref{eqphascurve}) does not minimise the BIC, making it useless in the global analysis. Afterwards we  test it again  with prior constraints from the global analysis without better success (Sect.~\ref{sec:planetaryatm}).  The MIPS light curves corresponding to the 14 different dithered positions are treated separately to take into account the spatial inhomogeneity of the detector response. Apart from a rescaling normalisation no baseline model is applied to the individual MIPS light curves.

Once we have selected the baseline models for all AORs, we perform a preliminary MCMC analysis (two chains of 80,000 steps) to assess the need for rescaling the photometric errors. For each AOR, the ratio of the residuals' standard deviation and mean photometric error is stored as $\beta_w$. This factor approximates the required rescaling of the white noise of each measurement. To take into account the correlation of the noise, a scaling factor $\beta_r$ is determined through the `time-averaging' method \citep{Pont2006} in which the scatter of individual points and the scatter of binned data with different time intervals are compared with the expected scatter of the corresponding binned data without the presence of correlated noise. The largest values of $\beta_r$ are kept.  At the end, the error bars are multiplied by the corresponding correction factor $CF = \beta_w \times \beta_r$. The values of $\beta_w$ and $\beta_r$ derived for each AOR are given in Table~\ref{tIRAC}. The $\beta_w$ values are generally higher than unity, indicating the undervaluation of photometric errors, e.g. the photometric IRAC errors on the mean are lower than the real photometric errors. Besides, light curves with $\beta_r >$ 1 highlight our inability of our models to perfectly represent the data (due to instrumental and/or astrophysical effects).
For the RVs, both our minimal baseline models correspond to a scalar  representing the systemic velocity of the star. To compensate both instrumental and astrophysical effects that are not included in the initial error calculation, the `jitter' noises of 3.8 (Keck/HIRES) and 1.7 m.s$^{-1}$ (HARPS)  are quadratically added to the error bars to equal the mean error to the standard deviation of the best-fit residuals.

\subsubsection{Global analysis}
\label{sec:globanalysis}

\begin{figure}[]
\centering
  \includegraphics[width=8.8cm,trim=0 0 0 20,clip=true]{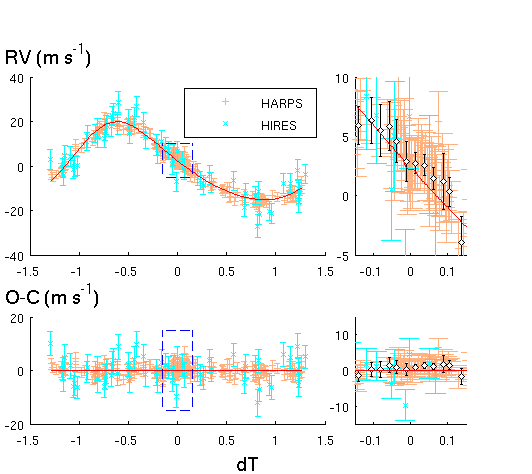}
  \caption{\label{rv} On the \textbf{top}, radial velocities measured with the HIRES and HARPS spectrographs. The data are period-folded on the best-fit transit ephemeris, the 0 of the {\it x-}axis corresponding to the inferior conjunction time. The fit is superimposed on red and the respective residuals are shown on the \textbf{bottom} panel. The left panels display those corresponding data during the whole planetary orbit and the right ones are a zoom on the box to highlight the (undetected) Rossiter-McLaughlin effect. The HARPS binned data are added in black diamonds filled white for clarity}
\end{figure}

Once the models selected and the errors rescaled, we perform a global MCMC analysis of our dataset to probe the posterior probability distributions of the jump and physical parameters. Two chains of 240,000 steps compose this analysis. We successfully check their good convergence and mixing with the statistical test of \citet{Gelman1992}, all jump parameters showing indeed a value close to 1 at the 0.01 level. The resulting median values and 68.3\% probability interval for the jump and physical parameters are given in Table~\ref{Oparam}. The fitted light curves are compared with the data  in Figs.~\ref{lcoccim}, \ref{lctra}, \ref{lcocc1}, and \ref{lcocc2}. Note that the light curves are not binned prior to the analysis (except for the subarray images, see Sect.~\ref{subsec:IRACdata}). The fitted radial velocities are displayed in Fig.~\ref{rv}. Unfortunately, the  Rossiter-McLaughlin effect is not detected. We also try to model the Rossiter-McLaughlin effect with the implementation of \cite{Boue2013} but results are not conclusive either. According to \cite{Winn2010}, we can expect to observe a maximum amplitude of the RV anomaly $\Delta V_{RM}\sim$4.10$^{-3}\times V\sin i \sim$12~m.s$^{-1}$ considering a maximum projected rotational velocity of 3 km.s$^{-1}$. However one should better expect $\Delta V_{RM}\sim$ 1.92~m.s$^{-1}$ assuming a stellar rotation of 48 days (see Sect.~\ref{subsec:RVresid}).

\begin{table}[!h]
\centering
\caption{\label{Oparam} System parameters derived from our global analysis of the {\it Spitzer} data and radial velocities.}
\begin{tabular}{lr}
\hline\hline
Parameter  & Value  \\ \noalign{\smallskip}
\hline  \noalign{\smallskip}
{\it Stellar parameters} & \\ \noalign{\smallskip}
  $\rho_* /\rho_{\odot}$    &          $  5.91_{-0.18}^{+0.17}  $ \\ \noalign{\smallskip}
  $M_*$ (M$_{\odot}$)     &            $  0.556 _{  -  0.065  }^{+0.071  }   $\\ \noalign{\smallskip}
{\it Orbital parameters} & \\ \noalign{\smallskip}
  $b = \frac{a \cos{i}}{R_{*}}$          &          $         0.7972 _{  -  0.0055 }^{  + 0.0053   }  $ \\ \noalign{\smallskip}
  $Duration$ (day) &          $                   0.04189  _{ -  0.00014 }^{  +  0.00014   }  $ \\ \noalign{\smallskip}
  $T_0 $      (BJD$_{TT}$) &          $        4865.084034 _{  -  0.000034  }^{ +  0.000035   }  $ \\ \noalign{\smallskip}
  $P $    (days)     &          $          2.64389803_{   -  0.00000025  }^{ +  0.00000027   }  $ \\ \noalign{\smallskip}
  $K $     (m/s)    &          $          17.59_{   -   0.25  }^{ +   0.25   }  $ \\ \noalign{\smallskip}
  $e $         &          $          0.1616_{   -  0.0032  }^{ +  0.0041  }  $ \\ \noalign{\smallskip}
  $\omega $ ($^{\circ}$) &          $            327.2_{   -   2.2  }^{ +   1.8  }  $ \\ \noalign{\smallskip}
  $a $         (AU) &          $          0.0308_{   -   0.0012 }^{  +   0.0013   }  $ \\ \noalign{\smallskip}
 $a/R_{*}$& $ 14.54   _{-   0.15}   ^{+   0.14}  $ \\ \noalign{\smallskip}
  $i $          ($^{\circ}$)&          $         86.858_{   -   0.052  }^{ +   0.049   }  $ \\ \noalign{\smallskip}
$\sqrt{V \sin (i)} \cos{\beta}$ ($\sqrt{m/s}$) &  $ -0.17 _{-0.52}^{+0.53}$ \\ \noalign{\smallskip}
$\sqrt{V \sin (i)} \sin{\beta}$ ($\sqrt{m/s}$) &  $ -0.09 _{-0.25}^{+0.26}$ \\ \noalign{\smallskip}
$\beta$ ($^{\circ}$) &  $ 181 _{-44}^{+43}$ \\ \noalign{\smallskip}
$V \sin (i)$ (km/s) &  $0.24 _{-0.17}^{+0.38}$ \\ \noalign{\smallskip}
{\it Planetary parameters} & \\ \noalign{\smallskip}
  $  dF $($\%$) & $                  0.6819   _{-  0.0028  }^{  +  0.0028   }  $ \\ \noalign{\smallskip}
  $  R_p (R_{\oplus})$ & $      4.10  _{ -   0.16  }^{  +   0.16  }  $ \\ \noalign{\smallskip}
  $  M_p (M_{\oplus})$ & $  25.4  _{ -   2.0  }^{  +   2.1   }  $ \\ \noalign{\smallskip}
{\it Occultation depths} (ppm) & \\ \noalign{\smallskip}
    3.6 $\mu$m & $        177. _{  -  43.  }^{  +  43.   }  $ \\ \noalign{\smallskip}
    4.5 $\mu$m & $       28. _{  -  18. }^{   +  25.   }  $ \\ \noalign{\smallskip}
    5.8 $\mu$m & $        229. _{  -  99.  }^{  +  107.   }  $ \\ \noalign{\smallskip}
    8 $\mu$m & $       362.  _{ -  29. }^{   +  29.   }  $ \\ \noalign{\smallskip}
    16 $\mu$m & $       1260.  _{ -  270.  }^{  +  280.   }  $ \\ \noalign{\smallskip}
    24 $\mu$m & $       1690. _{  -  460.  }^{  +  470.   }  $ \\ \noalign{\smallskip}
\hline  \noalign{\smallskip}
\end{tabular}
\end{table}

\begin{figure*}[]
\centering
  \includegraphics[width=18cm,trim=0 0 0 0,clip=true]{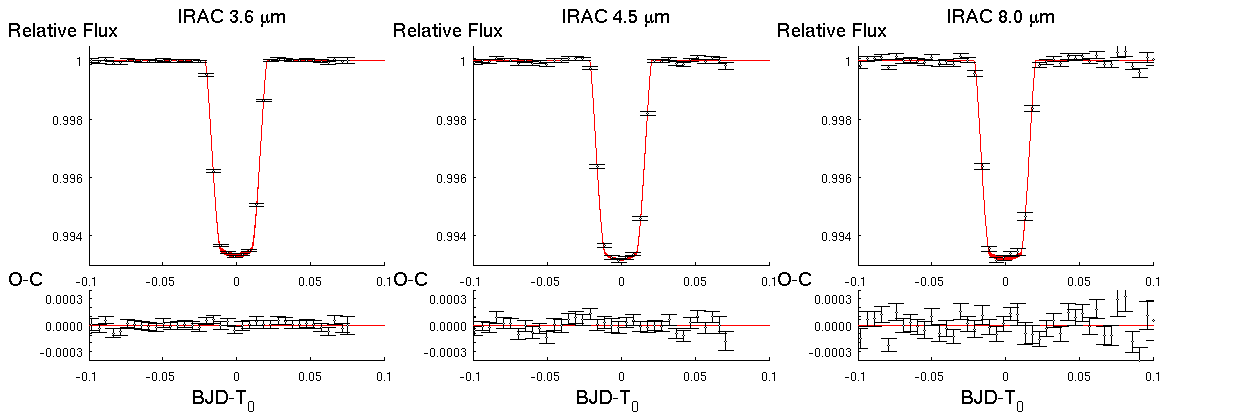}
  \includegraphics[width=18cm,trim=0 0 0 0,clip=true]{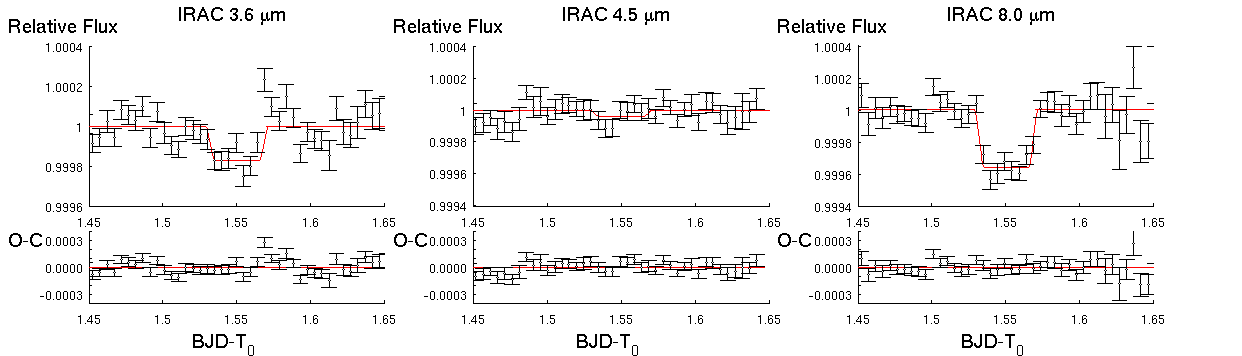}
   \caption{\label{combitraocc}  Detrended {\it Spitzer} 3.6-$\mu$m ({\it left}), 4.5-$\mu$m ({\it middle}), and 8-$\mu$m ({\it right}) photometry folded on the best-fit transit ephemeris obtained in our analysis per channel for the transits ({\it top}, Sect. \ref{sub:travari}) and in our global analysis for the occultations ({\it bottom}). The best-fit eclipse models are superimposed.}
\end{figure*}

From the inferred planet-star flux ratios we retrieve the brightness temperature $T_{bright}$ of the planet assuming the planet emits as a blackbody in the different {\it Spitzer} bandpasses and using a {\tt Phoenix} stellar atmosphere model \citep{Hauschildt1999} (with $T_{\textrm{eff}}=3400$~K, $\log g = 5.0$, and solar metallicity, Table~\ref{BrightnessTemp}). We warn the reader that the error bars do not take into account the error of the model. We finally derive the magnitude of the planet in the different IRAC bandpasses.

\begin{table}[]
\centering
\caption{\label{BrightnessTemp} Planetary brightness temperatures evaluated for every instrument/bandpass. The second column gives the planetary dayside temperature assuming a {\tt Phoenix} stellar atmosphere model with $T_{\textrm{eff}}=3400$~K.}
\begin{tabular}{cc}
\hline\hline
Wavelength	 & $T_{bright}$  \\ \noalign{\smallskip}
($\mu$m) & (K)  \\ \noalign{\smallskip}
\hline  \noalign{\smallskip}
3.6  & 922	$_{-54}^{+47}$ 		\\ \noalign{\smallskip}
4.5 & $\leq 690$ 			\\ \noalign{\smallskip}
5.6 & 716	$_{-96}^{+84} $		\\ \noalign{\smallskip}
8.0 & 683	$_{-19}^{+19}$ 	 \\ \noalign{\smallskip}
16.0 & 912$_{-125}^{+124}$ 		\\ \noalign{\smallskip}
24.0 & 1335$_{-294}^{+294}$		 \\ \noalign{\smallskip}
\hline  \noalign{\smallskip}
\end{tabular}
\end{table}

\begin{table}[]
\centering
\caption{\label{magnitude} Apparent and absolute magnitudes of GJ\,436 and its planet. We estimate the errors of the apparent and absolute magnitude around 0.07 mag and 0.12 mag respectively, taking into account the uncertainty in the absolute calibration, the photometric error, the uncertainty of the zero magnitude flux, and the error on the parallax.}
\begin{tabular}{c|cccc}
\hline\hline
 Wavelength	 & \multicolumn{4}{c}{Magnitudes} \\ \noalign{\smallskip}
 ($\mu$m)		& \multicolumn{2}{c}{Star} & \multicolumn{2}{c}{Planet}  \\ \noalign{\smallskip}
 & Apparent & Absolute & Apparent & Absolute \\ \noalign{\smallskip}
\hline  \noalign{\smallskip}
3.6  & 5.88 & 5.85 & 15.26 & 15.23  \\ \noalign{\smallskip}
4.5   & 5.84 & 5.81 & - & - \\ \noalign{\smallskip}
5.8  & 6.28& 6.25 & 15.38 & 15.35 \\ \noalign{\smallskip}
8.0   & 6.27 & 6.24 & 14.87 & 14.84 \\ \noalign{\smallskip}
\hline  \noalign{\smallskip}
\end{tabular}
\end{table}

\subsubsection{Analysis allowing for transit and occultation variations}
\label{sub:travari}

Our global analysis assumes a constant transit  depth and duration, orbital period, and occultation depth at every wavelength. Even if the resulting model fits very well our data,  some variations may indicate diverse physical phenomena which are relevant enough to be verified. For this purpose, we conduct new MCMC analyses, benefiting from the strong constraints brought by the global analyses to attain the highest sensitivity.

First we check if the transit depth $dF$ varies with wavelength. $dF$ provides the apparent radius of the planet, which gives in return a piece of the transmission spectrum of the terminator of the planet. We thus perform three separate additional MCMC analyses for each of the three relevant IRAC channels (3.6, 4.5 and 8 $\mu$m) considering only the light curves including a transit. We use a uniform prior distribution for $dF$ and assume normal prior distributions for the other jump parameters based on the posterior distributions resulting from  our main global analysis (Table~\ref{Oparam}).  We cannot detect any significant chromaticity of the transit depth (Table~\ref{varitratab}, Fig.~\ref{combitraocc}, and Sect.  \ref{sec:planetaryatm}).

\begin{table}[]
\centering
\caption{\label{varitratab} Transit depth variation with wavelength.}
\begin{tabular}{cc}
\hline\hline
Wavelength	 & Transit depth \\ \noalign{\smallskip}
($\mu$m) & (ppm) \\ \noalign{\smallskip}
\hline  \noalign{\smallskip}
3.6  & 6770	$\pm$ 42 \\ \noalign{\smallskip}
4.5 & 6881	$\pm$ 54 \\ \noalign{\smallskip}
8.0 & 6789	$\pm$ 61 \\ \noalign{\smallskip}
\hline  \noalign{\smallskip}
\end{tabular}
\end{table}

\begin{figure}
\centering
  \includegraphics[width=8.5cm]{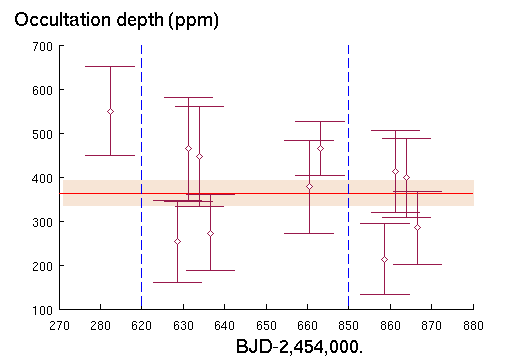}
  \caption{\label{saisoni4occ}  Occultation depth variation at 8 $\mu$m as function of time. Brown diamonds with their 1-$\sigma$-error bar represents occultation depths. The dashed blue lines indicate temporal caesura in the {\it x-}axis. The red horizontal line displays the eclipse depth value obtained during the global analysis and the shaded region surrounding it is its 1-$\sigma$-error bar.}
\end{figure}

\begin{figure}
\raggedright
  \includegraphics[width=5cm,trim=30 112 30 20,clip=true]{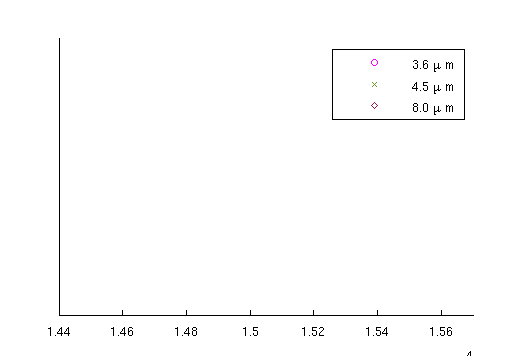}
\centering
  \includegraphics[width=8.8cm,trim=0 0 0 24,clip=true]{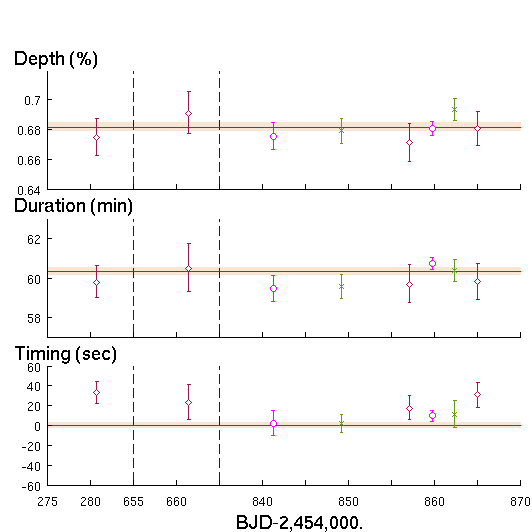}
  \caption{\label{varitra} Transit parameter variations. From top to bottom, the plots respectively show the transit depth, the transit duration, and the transit timing variation with their error bars according to time for the different IRAC channels. The pink dots stand for the 3.6-$\mu$m channel, the green crosses for the 4.5-$\mu$m channel and the brown diamonds for the 8-$\mu$m channel. The vertical dashed blue lines indicate temporal caesura in the {\it x-}axis. The horizontal red lines present in the two top panels respectively indicate the transit depth and duration derived from the global analysis. The shaded surrounding area is its 1-$\sigma$ error bar. }
\end{figure}

Then we test the transit and occultation depth variations as a function of time within a given instrument. Such transit depth variations could reveal a physical change of the radius with time (for e.g. due to the thermal expansion of the atmosphere or the presence of  variable clouds) or stellar activity (see \citetalias{Knutson2011} for all those aspects). Besides, the occultation depth fluctuation should provide constraints on general circulation models. We separate all light curves including an eclipse event in groups according to their channel origin and perform new MCMC analyses for each group separately. We set the transit and occultation depths as jump parameters for each individual eclipse and keep the other parameters as normal prior distributions based again on our global analysis (Table~\ref{Oparam}). We find no significant occultation depth variation (Table~\ref{occultationperCanal} and Fig.~\ref{saisoni4occ}  for the data at 8 $\mu$m), and no transit depth variation contrary to \citetalias{Beaulieu2011} and \citetalias{Knutson2011} studies (Fig.~\ref{varitra} and Table~\ref{TDV}). 

\begin{table}[]
\centering
\caption{\label{occultationperCanal} Individual measurements of the occultation depths for the different AORs (Sec. 3.4.3).}
\begin{tabular}{ccr}
\hline\hline
Wavelength & AOR  & Depth \\ \noalign{\smallskip}
($\mu m$) &    &    (ppm) \\ \noalign{\smallskip}
\hline  \noalign{\smallskip}
  $3.6 $    & 24882688 & $ 169.5 ^{  -  47.2  }_{ +  47.8}$\\ \noalign{\smallskip}
  $3.6 $    & 40848384& $   217.2 ^{  -  121.6 }_{  +  139.9}$\\ \noalign{\smallskip}
  $4.5 $    & 24882944& $   101.9  ^{ -  59.3 }_{  +  67.4}$\\ \noalign{\smallskip}
  $4.5 $    & 40848128& $   21.1 ^{  -  14.5 }_{  +  23.4}$\\ \noalign{\smallskip}
  $8.0 $    &23618304& $  549.9 ^{  -  100.7 }_{  +  100.7}$\\ \noalign{\smallskip}
  $8.0 $    &26812928& $  253.4 ^{  -  93.3  }_{ +  94.1 }$\\ \noalign{\smallskip}
  $8.0 $    &27604736& $  464.1 ^{  -  119.8  }_{ +  116.6}$\\ \noalign{\smallskip}
  $8.0 $    &27604992& $  447.2 ^{  -  114.8  }_{ +  113.3}$\\ \noalign{\smallskip}
  $8.0 $    &27605248& $  272.7 ^{  -  85.9  }_{ +  88.3}$\\ \noalign{\smallskip}
  $8.0$    &27863296& $  378.0  ^{ -  105.4  }_{ +  105.9}$\\ \noalign{\smallskip}
  $8.0 $    &27863808& $  464.8 ^{  -  61.7  }_{ +  62.3}$\\ \noalign{\smallskip}
  $8.0 $    &28970240& $  285.2 ^{  -  84.5  }_{ +  82.5}$\\ \noalign{\smallskip}
  $8.0$    &28969472& $  211.6 ^{  -  78.6  }_{ +  82.6}$\\ \noalign{\smallskip}
  $8.0 $    &28969728& $  413.4 ^{  -  94.0  }_{ +  93.2}$\\ \noalign{\smallskip}
  $8.0$    &28969984& $  397.8 ^{  -  90.5  }_{ +  90.6}$\\ \noalign{\smallskip}
\hline  \noalign{\smallskip}
\end{tabular}
\end{table}

The transit duration variation, which might be due to the presence of another orbiting body such as a moon (e.g.,  \citealt{Kipping2009}), is the third parameter we examine. Because it might be linked to a transit depth variation, we set those two parameters as jump parameters in our MCMC analyses and maintain normal prior distributions for the other parameters. We perform individual MCMC analyses on each time-series recording a transit. We do not detect any significant transit duration variation (Fig.~\ref{varitra} and Table~\ref{TDV}).

\begin{table*}[t]
\centering
\caption{\label{TDV} Individual transit parameters. Columns 3 to 5 derive from our analysis. Our transit depths are compared with former studies in the last two columns.}
\begin{tabular}{cccccccc}
\hline\hline
Wavelength & AOR  & Depth  & Duration & Timing  & Depth from \citetalias{Beaulieu2011} & Depth from \citetalias{Knutson2011}\\ \noalign{\smallskip}
($\mu m$) &   &  ($\%$) & (day) & BJD$_{TT}-2450000. $&  ($\%$)&  ($\%$)   \\ \noalign{\smallskip}
\hline  \noalign{\smallskip}
  3.6     &    28894208 & $      0.675 ^{  -  0.009 }_{  +  0.009 }    $  & $0.0413   ^{-  0.0004}_{  +  0.0005} $& $4841.28898 ^{ -0.00014  }_{ +0.00015}$ & & $0.669 \pm 0.006$  \\ \noalign{\smallskip}
  $3.6 $    & 28894464 & $ 0.680  ^{  -  0.004 }_{    +  0.004 }    $  & $  0.0422  ^{ -  0.0002  }_{   +  0.0002}    $  &$4859.79635  ^{-0.00006  }_{ +0.00006} $ & $0.712 \pm0.006$&$0.722 \pm0.010$  \\ \noalign{\smallskip}
  $4.5 $    & 28894720 & $ 0.678   ^{ -  0.009 }_{    +  0.008 }    $  & $  0.0414   ^{ -  0.0004  }_{   +  0.0004}    $  &$4849.22066 ^{ -0.00010  }_{+0.00011}$ &$0.638\pm 0.018$&$0.687\pm 0.008$\\ \noalign{\smallskip}
  $4.5 $    & 28894976 & $ 0.693   ^{ -  0.007 }_{    +  0.007 }    $  & $  0.0419   ^{ -  0.0003  }_{   +  0.0004}    $  & $4862.44027 ^{ -0.00015 }_{+0.00015}$ & &$0.723\pm 0.010$ \\ \noalign{\smallskip}
  $8.0 $    & 23515648 & $ 0.674   ^{ -  0.012  }_{   +  0.012  }    $  & $ 0.0415   ^{ -  0.0006 }_{   +  0.0006}    $  & $4280.78296 ^{  -0.00013  }_{  +0.00013 }$ &$0.685\pm 0.012$&$0.693\pm 0.009$\\ \noalign{\smallskip}
  $8.0 $    & 27863552 & $ 0.690   ^{ -  0.013  }_{   +  0.014 }    $  & $  0.0420    ^{-  0.0008  }_{   +  0.0009}    $  & $ 4661.50416 ^{ -0.00021 }_{  +0.00021 }$&$0.675\pm 0.012$&$ 0.680\pm 0.010$\\ \noalign{\smallskip}
  $8.0 $    & 28895232 & $ 0.670   ^{ -  0.012 }_{   +  0.013 }    $  & $  0.0414   ^{ -  0.0006  }_{   +  0.0007}    $  & $ 4857.15254 ^{ -0.00014  }_{ +0.00014 }$&$0.715\pm 0.013$&$ 0.676\pm 0.008$\\ \noalign{\smallskip}
  $8.0 $    & 28895488 & $ 0.680   ^{ -  0.011  }_{   +  0.011 }    $  & $  0.0415    ^{-  0.0006 }_{    +  0.0007}    $  & $4865.08439 ^{ -0.00015 }_{  +0.00015}$&&$0.710\pm 0.008$ \\ \noalign{\smallskip}
\hline  \noalign{\smallskip}
\end{tabular}
\end{table*}

Finally, we search for transit timing variations (TTV), which may disclose the presence of another orbiting body in the system \citep[][and references therein]{Holman2005,Agol2005}. We perform a new individual analysis of the transit light curves, keeping all the parameters except the mid-transit times under the control of Bayesian penalties based on the results of our global analysis. The measured timings are given in Table~\ref{TDV}. When compared with the transit ephemeris deduced in our global analysis, they do not reveal any significant TTV (Fig.~\ref{varitra}, bottom panel).

\subsubsection{Flares-like structures in the light curves}
\label{sec:dataa_flare}

\begin{figure}
\centering
  \includegraphics[width=8.8cm]{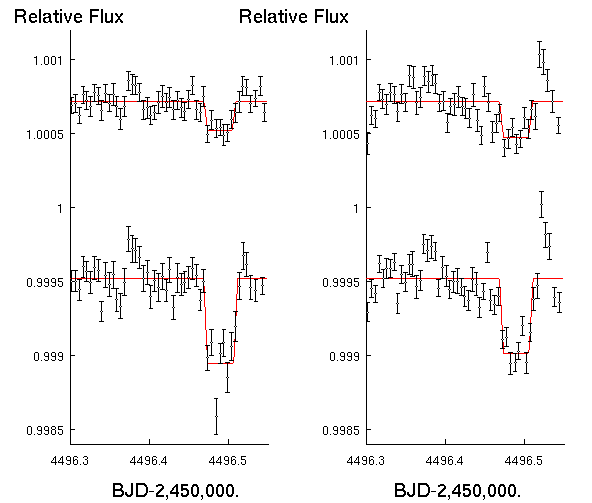}
  \caption{\label{flarefwhm}  Influence of the chosen aperture radius and the baseline modelling. We present the same light curve (at 3.6 $\mu$m, 2008 January 30, AOR: 24882688) four times, however showing different noise level and a different occultation depth due to the data reduction and analysis. On the {\bfseries left}, the aperture photometry is centred by a 2D elliptical Gaussian fit with an aperture radius of 1.9 pixels, while on the {\bfseries right} the aperture radius equals 3.5 pixels. They all proceed to the same MCMC analysis with the inferred parameters from Table~\ref{Oparam}, except that the {\bfseries top} ones contain a baseline model dealing with the measured PSF FWHM, while the \textbf{bottom} ones do not. 
 }
\end{figure}

Previous studies of GJ\,436 {\it Spitzer} datasets reported structures that could potentially be attributed to stellar flares. The first one was signalled just after an occultation at 3.6 $\mu$m (2008 January 30, AOR : 24882688) by \citetalias{Stevenson2010} and \citetalias{Beaulieu2011} and led to contradictory results.  Their opinions diverged on how to take care of the post occultation spike they both identified. On our side, we notice that the amplitude of the structure highly depends on the chosen aperture radius, and disappears for aperture radii below 2.1 pixels (Fig.~\ref{flarefwhm}). 
The insertion of  Eq.~\ref{eqfwhm} in the baseline model leads to a significantly decreased occultation depth ($\sim$170 ppm instead of $\sim$535 ppm) and to a higher photometric accuracy  (mean photometric error $\sim$ 5.11 10$^{-4}$ instead of 10.30 10$^{-4}$, 
 Fig.~\ref{flarefwhm}), which we assign to the strong evolution of the FWHMs during the occultation. Thanks to the inclusion in our modelling of  the impact of the FWHM variations, a series of tests reveal that our measured occultation depth of $169\pm48$ ppm is not dependent on the selected aperture radius or on the rejection of a fraction of the post-occultation data, demonstrating its robustness. 

The second  occultation time-series at 3.6 $\mu$m was supposed to confirm or infirm \citetalias{Stevenson2010} and \citetalias{Beaulieu2011} interpretations but another flare-like structure was recorded during the occultation by \citetalias{Stevenson2012} (see Fig.~\ref{lcocc1}, second light curve from the top, 2011 February 1, AOR : 40848384). In this case, we cannot identify any sharp variation of an external parameter able to explain it. We try to measure the occultation depth when discarding (or setting a zero-weight on) the last flux measurements that show a net discrepancy but the occultation depth cannot be constrained. So we cannot discuss the influence of this flare-like structure on the occultation depth measurement. The addition of this light curve to our global analysis cannot give a strong confirmation (217 $\pm$ 140 ppm) of our first 3.6~$\mu$m occultation depth measurement but agrees with it. 

\begin{figure}
\centering
  \includegraphics[width=8.8cm,trim=0 85 0 15,clip=true]{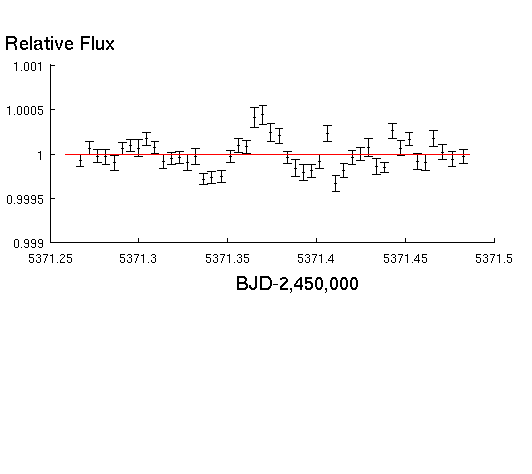}
  \caption{\label{lcflare}  Flux modulations present in the 2010 June 6 {\it Spitzer} light curve (AOR : 38807296) during GJ\,436 monitoring. }
\end{figure}

\begin{figure}
\centering
  \includegraphics[width=8.8cm,trim=20 10 20 285,clip=true]{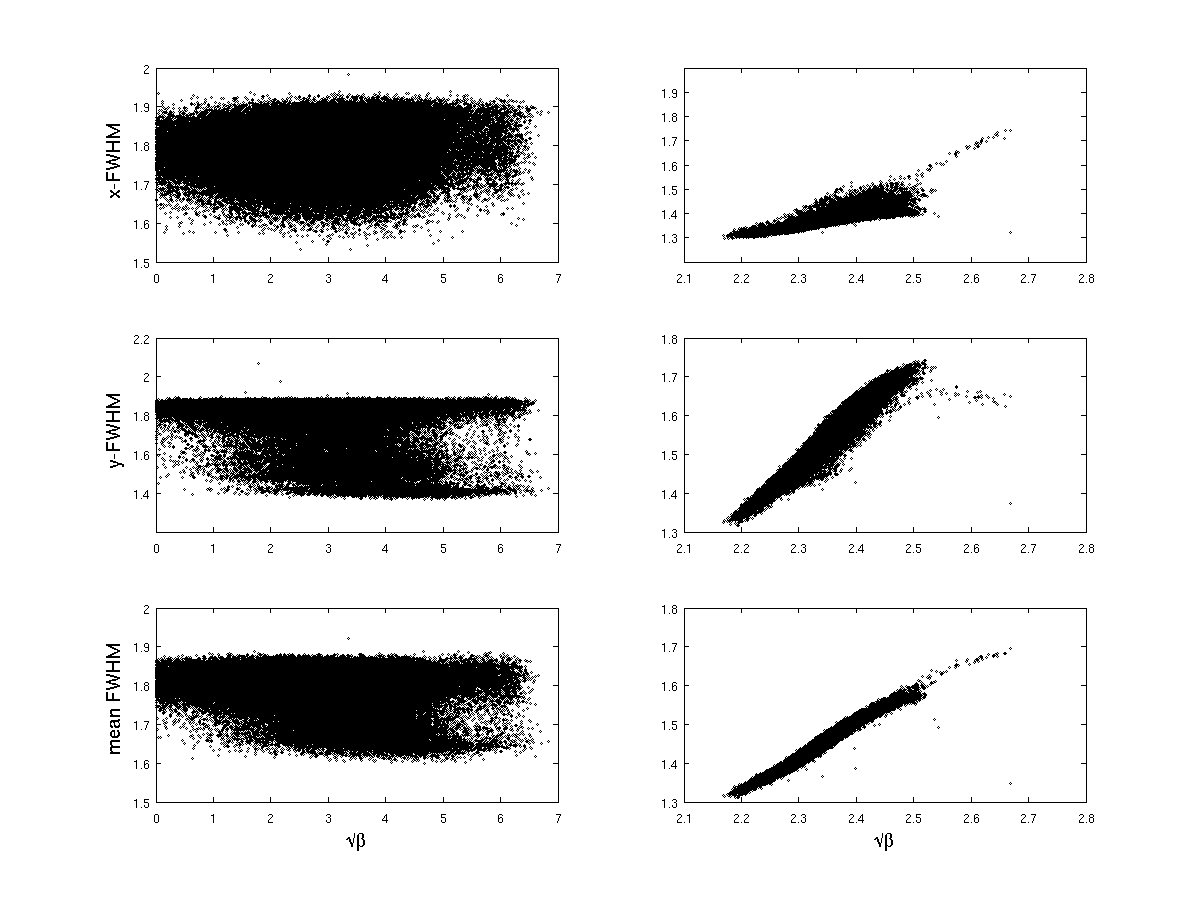}
  \caption{\label{fwhmbetaflare}  Lack of correlation on the PSF FWHM of GJ\,436 (AOR : 38807296) with the square root of the pixel noise parameter ($\sqrt{\beta}$) on the left. For comparison, the same graph corresponding to a light curve from the same bandpass (AOR : 28894464) is displayed on the right. The mean FWHM is clearly proportional to the square root of the pixel noise parameter, as expected.}
\end{figure}

We finally report another flare-like structure during an observation of GJ\,436 at 3.6~$\mu$m. This one happened outside a scheduled eclipse (2010 June 23, AOR : 38807296, Fig.~\ref{lcflare}). We cannot remove this peak with any of our baseline models. However, for that AOR we notice a surprising absence of correlation between the noise pixel parameter and the measured PSF FWHM (Fig.~\ref{fwhmbetaflare}), suggesting again an instrumental origin of the photometric structure. In the end, we can only identify one flare-like structure of which the origin is probably instrumental (cosmic hit in the PSF core), considering the absence of other similar structures in our extensive data set and the extreme quietness of GJ\,436. Indeed, the S$_{HK}$ index, a proxy for stellar activity, measured from the HARPS spectra is weak in comparison to the Ca~{\sc ii} H\&K emission of similar type stars  and does not significantly vary on short time scales (Astudillo et al. in prep). 

\section{Planetary companion}

In this section, we present the search for a second planet orbiting GJ\,436, which we perform on the residual light curves and the HARPS RVs. We also discuss the two potential transiting companions found recently by \citetalias{Stevenson2012} in some {\it Spitzer} datasets. 

\subsection{Search in the residual light curves}
\label{subsec:marplec}

We analyse the best-fit residual light curves resulting from our global MCMC analysis in order to be detached from GJ\,436b eclipses and from any instrumental systematics. 
We use our own version of the algorithm MISS MarPLE \citep{BertaMarple2012} on the residual light curves. 

\begin{figure}
\includegraphics[width=8.8cm,trim=0 0 0 25,clip=true]{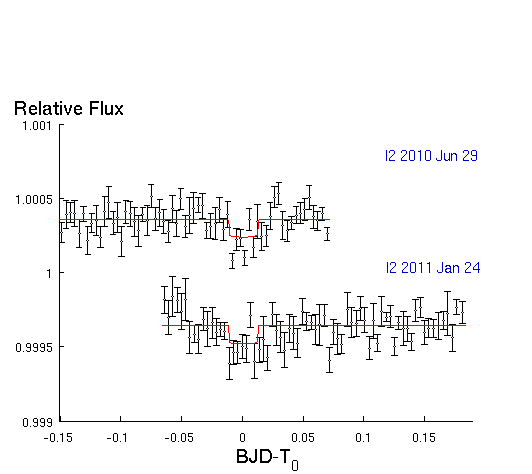} 
\caption{\label{lcminec}  Transit-like structures detected in two residual light curves by our own version of the code MISS MarPLE  \citep{BertaMarple2012}. For both structures, a box-like model illustrates a transit depth of 120~ppm and a duration of 0.6~h.}
\end{figure}

We can identify only two potential transit-like events. The strongest one corresponds to a transit depth of $\sim$150 ppm lasting for 0.5 hour at BJD = 2,455,376.702 with a $\sim$4-$\sigma$ significance. The second one recovers a signal of  only a 3.1-$\sigma$ significance at BJD = 2,455,585.686 and lasts for 0.6~hour with a depth of $\sim$100~ppm. It should be noted that the correlated noise is not fully taken into account when estimating those significance levels. The actual significance of both structures is clearly marginal, as can be judged by eye in Fig.~\ref{lcminec}. With 1475~measurements from both light curves and 4 more degrees of freedom for the transit model, a $\Delta \chi^{2}$~=~-6 between the transiting and non-transiting companion model corresponds to a $\Delta$BIC~=~-6~+~4~log(1475)~=~+6.7. Using the BIC as a proxy for the model marginal likelihood \citep{Kass1995}, this $\Delta$BIC results in an approximated Bayes factor of $e^{\Delta BIC/2} \sim$~28   in favour of the non-transiting companion model. 

These two transit-like features (Fig.~\ref{lcminec}) were also detected by \citetalias{Stevenson2012} although with a higher depth and longer duration. These authors attributed them to their planet candidate UCF-1.01.

\begin{figure}
\includegraphics[width=8.8cm,trim=0 0 0 0,clip=true]{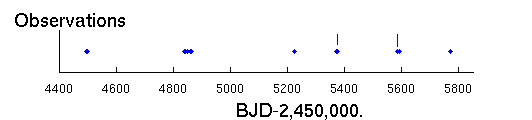} 
\includegraphics[width=8.8cm,trim=0 0 0 25,clip=true]{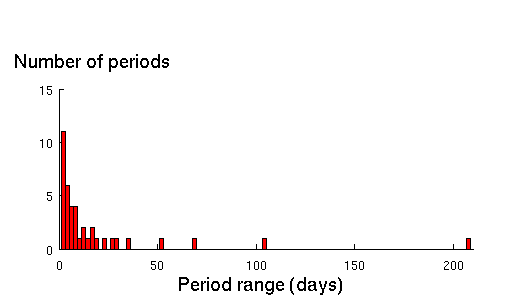} 
\caption{\label{fig:perposs}  {\bf Top:} Timing of GJ\,436 observations with the 3.6 and 4.5 $\mu$m IRAC channels. The two vertical lines indicate the datasets having the  transit-like features. {\bf Bottom:} Distribution of the possible period range for the putative additive transiting planet. The observational coverage discards all periods below 1.18~days.}
\end{figure}

Assuming that both structures correspond to two transits of the same and still undetected planet, the elapsed time between the two signals gives a maximum period of $\sim$ 208.98 days while the transit duration regarding a central transit returns a minimal period of 0.098 day. The lack of a continuous monitoring between these two potential transits prevents us from identifying a single period. Still, our extensive dataset allows us to discard many fractions of 208.98 days for which at least a third transit should have been present in the {\it Spitzer} data. Injecting fake transits at the corresponding phases, we check our ability to recover them in the {\it Spitzer} data with MISS MarPLE for each period and each expected transit. Fig.~\ref{fig:perposs} shows the period ranges that we can keep from these tests. In this figure, the periods too close to GJ\,436b's one are also discarded to allow the dynamical stability of the system \citep[from $\sim$1.8 to $\sim$3.9 days,][]{Ballard2010TT}. In the end 40 possible periods remain. One may assess their replicability with new ultra-precise time-series photometry based on the ephemeris  $2455476.702 + N \times 208.98$ days. 
For the sake of completeness, we also analyse the residual light curves with two versions of the algorithm BLS \citep{Kovacs2002} (the Optimal BLS from \citet{Ofir2013} and  the BLS available on the NASA Exoplanet Archive website\footnote{http://exoplanetarchive.ipac.caltech.edu}), which searches for periodic box-shaped structures in photometric time-series. We fail to detect any significant signal.

\subsection{Discussion on a putative GJ\,436c}

\citetalias{Stevenson2012} proposed two planetary candidates in the GJ\,436's stellar system and named them temporarily UCF-1.01 and UCF-1.02 until their confirmation. They observed UCF-1.01 transits during 2008 July 14, 2010 January 28, 2010 June 29, 2011 January 24 and July 30 datasets. UCF-1.02 potential transits occurred during both of those 2010 datasets. 

\begin{figure}
\includegraphics[width=8.8cm]{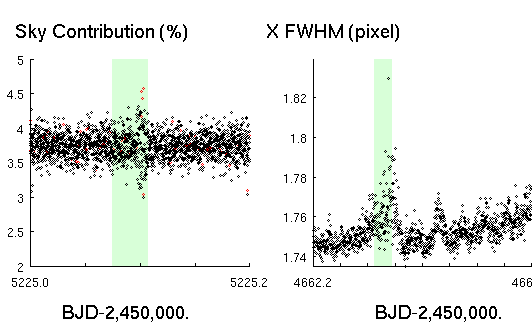} 
\caption{\label{probfit}  Atypical behaviour of measured parameters during two of \citetalias{Stevenson2012}'s UCF-1.01 transits. The figure on the \textbf{left} shows a background contribution discontinuity around BJD 2,455,225.09. The one on the \textbf{right} points out a strong fluctuation of the {\it x}-FWHM around BJD 2,454,662.32. UCF-1.01 transit events are shaded in green.}
\end{figure}

In our analysis we pay particular attention to the behaviour of the measured external parameters. The following two examples may cast doubt on some observations of \citetalias{Stevenson2012}'s candidate transits. A change of the background contribution (Fig.~\ref{probfit}) indeed affects the 2010 January 28 dataset  just at the end of the observation of UCF-1.01 transit, while a larger {\it x}-FWHM alters the 2008 July 14 light curve   during another UCF-1.01 transit. \cite{Ballard2010b} also discarded the latter example because only small photometric apertures reveal a transit-like shape.

The signal-to-noise quality of our residual light curves being sufficient to identify transit shapes similar to \citetalias{Stevenson2012}, we must also verify that our adopted baseline models are not responsible for the subtraction of the transit signals.
We thus insert the transit signals observed by \citetalias{Stevenson2012} in our original light curves and perform individual MCMC to get new residual light curves. We look for transit-box shapes on the individual residual light curves with the previous manner and identify the injected transits with more than 4-$\sigma$ confidence. Our models correcting for instrumental systematics do not destroy transit signals.

\begin{figure}
\includegraphics[width=8.8cm,trim=0 0 0 45,clip=true]{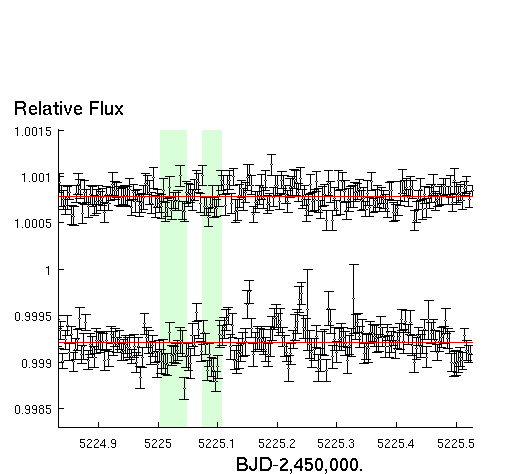} 
\caption{\label{compac}  Highlight on the impact of the data reduction/analysis details on the presence of low-amplitude transit-shaped structures in the {\it Spitzer} photometry (2010 Jan 28, AOR: 38702848). We apply relative flux offsets in the above two plots for clarity. The upper dataset is the result of aperture photometry with a 2D elliptical Gaussian centring while the lower one is done with a double 1D Gaussian fit. The upper is corrected with a baseline involving the measured FWHM while the lower is not. Black dots are the binned data per interval of 5 min with 1-$\sigma$ error bars. The supposed UCF-1.01 (right) and UCF-1.02 (left) transit events are shaded in green. }
\end{figure}

\begin{figure}
\includegraphics[width=8.8cm,trim=0 0 0 45,clip=true]{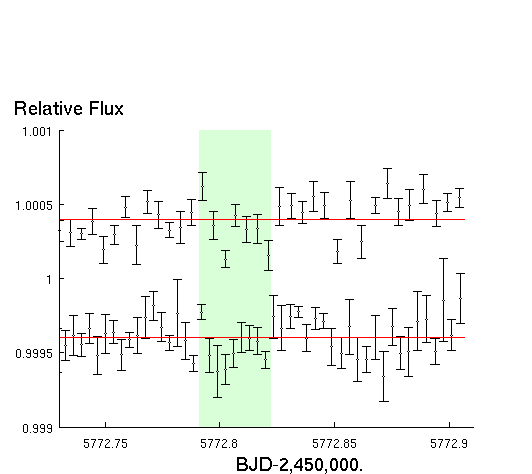} 
\caption{\label{compai2h3}  Same as Fig.~\ref{compac} for the 2011 Jul 30 data (AOR:  42614016). The upper dataset is the result of aperture photometry with a 2D elliptical Gaussian centring and is corrected for all systematics minimising the BIC. The lower dataset is produced from a double 1D Gaussian fit of the computed {\it x-}FWHM and corrected in the same way.  The shaded green area corresponds to UCF-1.01 transit event.}
\end{figure}

We wonder why our results diverge. We both opted for a similar aperture radius (\citetalias{Stevenson2012} : 2.25 px; our study : 2.2 px) and for a BLISS mapping in the 2010 January 28 dataset, but we differ in the choice of the centring technique. \citetalias{Stevenson2012} employed their time-series image de-noising for this purpose before centring with a Gaussian whereas we use a 2D elliptical Gaussian fit. We can obtain a similar structure to \citetalias{Stevenson2012} (Fig.~\ref{compac}), if we centre the PSF with a double 1D  Gaussian having the computed {\it x-}FWHM  and do not model the light curve with the FWHMs (Eq. \ref{eqfwhm}), but the UCF-1.01 transit signal-to-noise ratio remains very low.
Likewise, a similar transit-like shape in the 2011 July 30 light curve is found using the same aperture radius as \citetalias{Stevenson2012} (5 px) with a PSF centring obtained by a double 1D Gaussian fit (Fig.~\ref{compai2h3}) and with the same baseline as the one from our analysis. However, the transit is not significant ($\sim$1 $\sigma$).

\subsection{Analysis from the radial velocity residuals}
\label{subsec:RVresid}

\begin{figure}
\includegraphics[width=8.8cm,trim=0 0 0 0,clip=true]{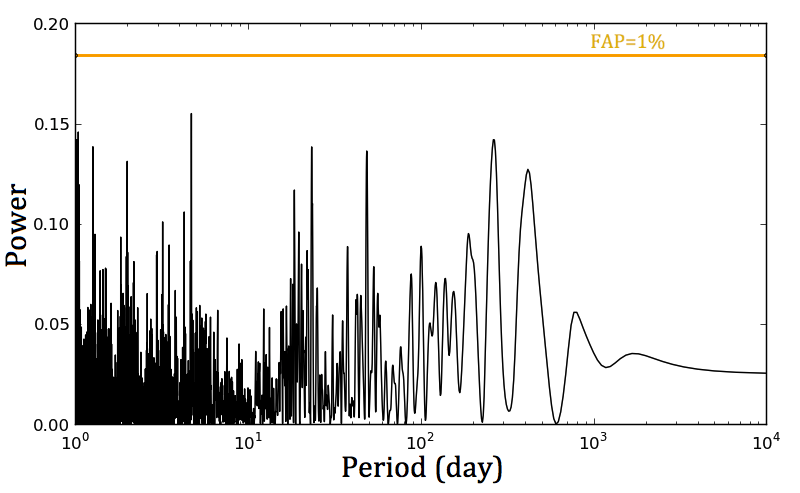} 
\caption{\label{periodo}  GJ\,436 periodogram of the residuals for the model composed of 1 planet on a Keplerian orbit plus 1 drift. The horizontal yellow line represents the 1\% false alarm probability. No peak shows a significant power excess. }
\end{figure}

\begin{figure}
\includegraphics[width=8.8cm,trim=0 0 0 0,clip=true]{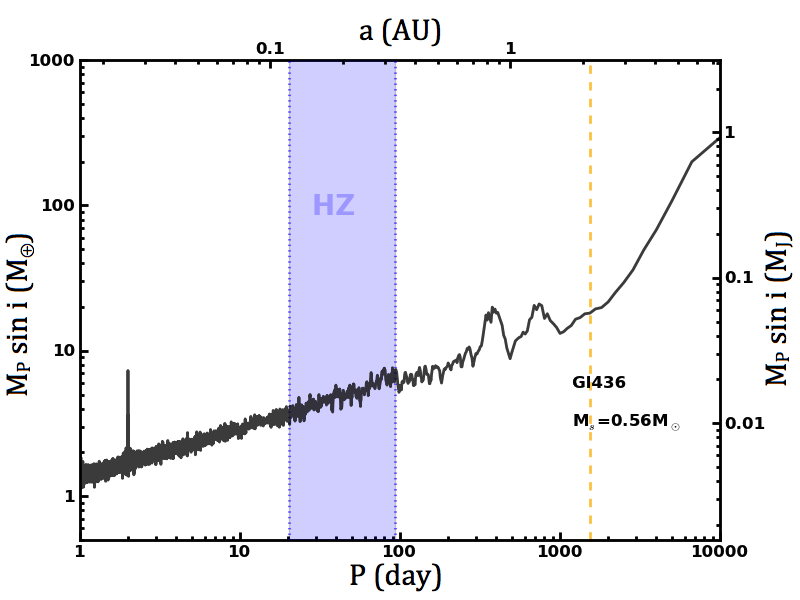} 
\caption{\label{dlm2}  Conservative detection limits on $M_{p} \sin i$ applied to GJ\,436 residual RV time-series around a chosen model (composed of planets, linear drifts, and/or a simple sine function). Planets above the limit are excluded, with a 99\% confidence level, for all 12 trial phases. The vertical dashed yellow line marks the duration of the survey.}
\end{figure}

The standard deviation for the RV residuals around a Keplerian solution composed of a single planet is 1.53 m/s, its mean error of 0.02 m/s, and the resulting $\chi^2$ of 2.0$\pm$0.2. In their periodogram corresponding to the one planet + drift model (Fig.~\ref{periodo}), we identify several peaks although none shows a significant power excess. Eight peaks have power excess above the 1-$\sigma$-significance level (which corresponds to a power p=0.131). They have periods 1.0181, 1.0417, 1.266, 2.000, 4.70, 23.4, 48.8 and 263 days with powers 0.142, 0.146, 0.131, 0.155, 0.138, 0.136 and 0.142, respectively. We note that one is at about twice the period of GJ\,436b, at 4.7 days. We also identify a peak around 48 days, which is reminiscent of the rotational period identified by \citet{Demory2007} (which differs from the rotational period identified by photometry by \citetalias{Knutson2011}, P$\sim$57 days) based on spectral indices for a sub-sample of our HARPS spectra, and another power excess around 23 days, i.e. about half the 48-day period and thus possibly a harmonic. Nevertheless, they all have a power much below the power threshold for the 3-$\sigma$-confidence level ($\mathcal{P}$=0.187) and we therefore cannot draw any firm conclusion. Assuming that the residuals are noise only, Fig.~\ref{dlm2} shows the detection limit, which, for a given period, delineates the mass limit above which a planet on a circular orbit is excluded (with a confidence level of 99\% and for 12 randomly chosen trial phases; see \citealt{Bonfils2013}). We see that for specific periods, we can exclude planets with masses higher than Earth-mass and reject planets with masses above 10 Earth masses for periods up to few-hundred days. We exclude super-Earths more massive than 3-5 Earth masses in GJ\,436's habitable zone, which is spread from 0.121 and 0.330 AU basing on \cite{Selsis2007} criteria. The analysis of the RV residuals cannot constrain the presence of UCF1.01 and UCF1.02 given that their small radii ($\sim$0.7~R$_{\oplus}$) should correspond to masses lower than our upper-limit mass.

\section{Atmospheric analysis}
\label{sec:planetaryatm}

In order to use our new measurements to draw inferences on the atmospheric properties of GJ\,436b we employ a number of atmospheric modelling tools.  We first derive planetary pressure-temperature \emph{P--T} profiles using the 1D plane-parallel model atmosphere code described in \citet{Fortney05,Fortney08a}.  The opacity database is described in \citet{Freedman08} and the equilibrium chemistry calculations in \citet{Lodders02}.  We use the base solar abundances of \citet{Lodders03}.  

\begin{figure}
\centering
\includegraphics[width=8.8cm,trim=0 0 0 0,clip=true]{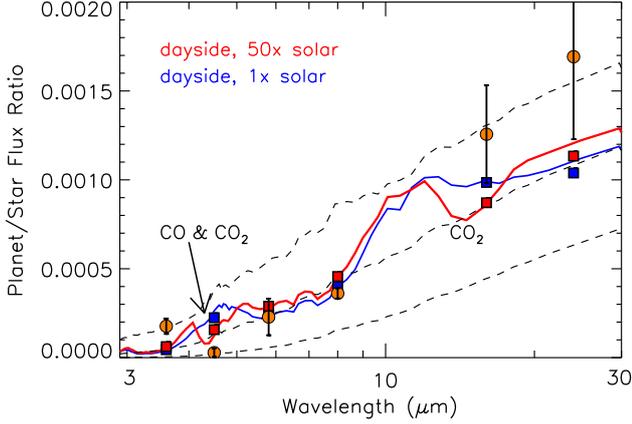}
\caption{Model planet-to-star flux ratios compared to the observed \emph{Spitzer} occultation data.  Both models assume a hot day side, with absorbed energy redistributed over the dayside only.  The blue model uses solar metallicity, while the red model uses 50$\times$-solar.  The \emph{Spitzer} data are orange circle while the model band averages in the six bands are shown as coloured squares.  The location of prominent absorption features that differ between the models (where CO and CO$_2$ absorb strongly) are labelled.  The planet-to-star flux ratios of 500, 700, and 900~K blackbodies are shown in dashed black.
\label{ratio}}
\end{figure}

We first generate thermal emission spectra for models representative of conditions on the dayside of the planet.  Within the framework of the thermal emission calculations, we are constrained to pre-tabulated equilibrium chemistry abundances.   In Fig.~\ref{ratio} we plot two models.  One uses solar abundances and the other 50$\times$-solar, broadly similar to the carbon abundance in Uranus and Neptune \citep{Baines95}.  At higher metallicity, hydrogen-poor molecules such as CO and CO$_2$ become more abundant \citep{Lodders02}, with the mixing ratio of CO rising linearly, and CO2 quadratically, with metallicity.  While this metal-enhanced model certainly goes in the right direction towards reproducing the large measured flux ratio differences between the 3.6 and 4.5 $\mu$m IRAC bandpasses, the fit is not satisfactory and our occultation depth at 3.6 $\mu$m, which is smaller than \citetalias{Stevenson2010}'s, is still too high to be explained by a CH$_4$ fluorescence with only stellar photons \citep{Waldmann2012}.
 \citetalias{Stevenson2010} and \citet{Madhu11a} have previously shown that extraordinarily low CH$_4$ mixing ratios, along with high CO and/or CO$_2$ abundances are needed to reproduce the previous \citetalias{Stevenson2010} observations.  We concur on this point based on our measurements.  Inversion techniques could be used on to better constrain the atmospheric abundances \citep[e.g.,][]{Madhu11a,Line12}. 
 
\begin{figure}
\centering
\includegraphics[width=8.8cm,trim=0 0 0 0,clip=true]{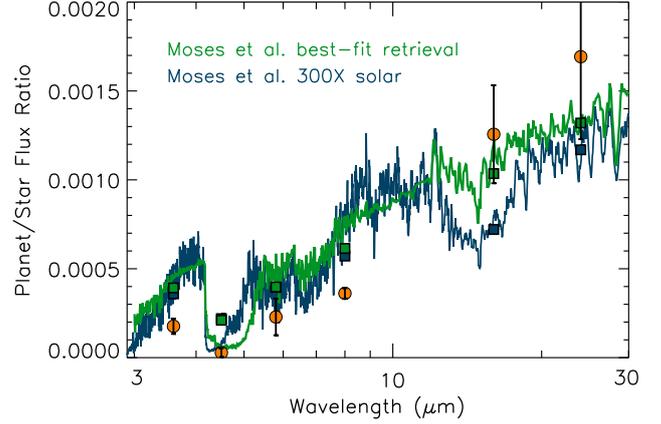}
\caption{Comparison of our observed \emph{Spitzer} occultation data with synthetic emission spectra for GJ\,436b from \citet{Moses2013}. The green and blue curves correspond respectively to the best fit retrieved model, given \citetalias{Stevenson2010} data, using the method of \citeauthor{Line2013} and the 300$\times$-solar model. The model band averages in the six bands are shown as coloured squares. Our \emph{Spitzer} data are shown as orange circles.
\label{fig:line}}
\end{figure}

In Figure~\ref{fig:line} we compare our \emph{Spitzer} depths to two models from \cite{Moses2013} that were created to better fit the previous \emph{Spitzer} depths derived by \citetalias{Stevenson2010}.  The model labelled ``best-fit retrieval'' used the Bayesian inversion methods of \cite{Line2013,Line2014} to find a best-fit spectrum, with free parameters that include the $P-T$ profile and chemical mixing ratios.  The ``300$\times$-solar'' model uses a \citeauthor{Line2013} $P-T$ profile, but mixing ratios from a detailed non-equilibrium chemistry calculation for an atmosphere enriched in metals by a factor of 300 over solar abundances.  It is clear that reduced secondary eclipse depths at 3.6, 5.8, and 8.0 $\mu$m from our analysis lead to a poor fit from these models.  Given these lower fluxes, it appears plausible that a cooler atmospheric $P-T$ profile would better fit these data, and also yield a lower flux in the 4.5 $\mu$m band, which has always been problematic to fit \citep[\citetalias{Stevenson2010};][]{Madhu11a,Moses2013}.  It does appear that some CH$_4$ depletion and CO/CO$_2$ enhancement will still be needed to match 3.6 and 4.5 $\mu$m band depths, given our models presented in the previous plot (Fig.~\ref{ratio}).  Also, cooler day-side temperatures may allow for a fit with a lower metallicity, and hence less CO$_2$, and a better fit at 16~$\mu$m, which includes a strong CO$_2$ band.

\begin{figure*}
\sidecaption
\includegraphics[width=12cm,trim=0 0 0 0,clip=true]{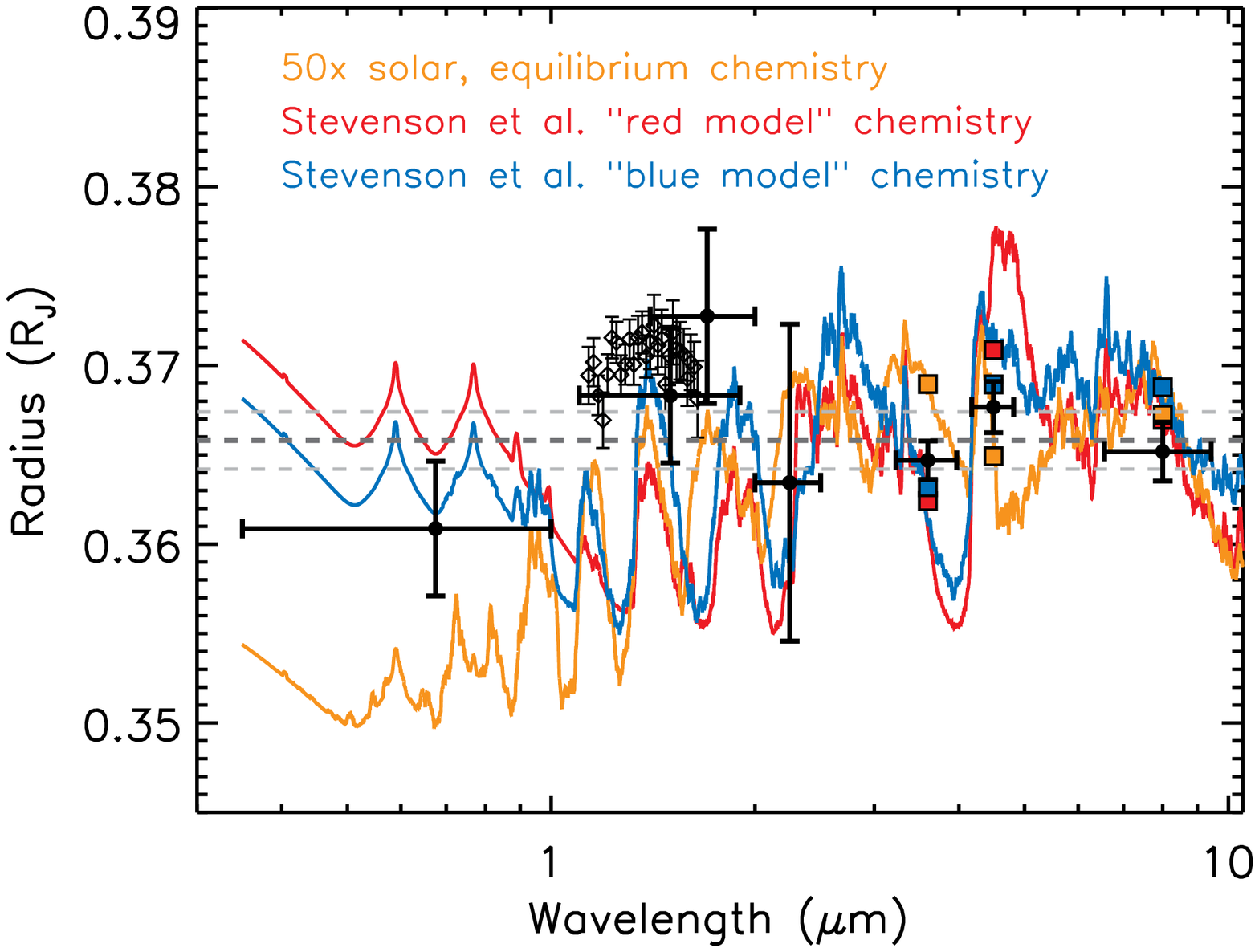}
\caption{\label{transit} Model transit radius as function of wavelength compared with published transit depths and our \emph{Spitzer} results. From left to right, the radii are derived from EPOXI \citep{Ballard2010TT},  HST WFC3 \citep[][represented by empty diamonds with thinner error bars in the 1.14-1.65~$\mu$m wavelength range]{Knutson2014}, HST NICMOS \citep{Pont2009}, the {\it H} and {\it K} bands from the ground \citep{Alonso2008,Caceres2009}, and IRAC (ours) studies. Three models are shown.  In orange is the 50$\times$-solar model from Fig.~\ref{ratio}.  In red and blue are transmission spectra taken from \citet{Shabram11}, who used mixing ratios suggested for the planetary dayside by \citetalias{Stevenson2010}.  Band-average calculations are shown as squares for the three models, across the \emph{Spitzer} bandpasses.  The dramatic differences between the models at optical wavelength are only due to different assumptions about the abundances of alkali metals.  The blue and red models, which have depleted CH$_4$ and enhanced CO and CO$_2$, are generally preferred.}
\end{figure*}

We next turn to the transmission spectrum of the planet to examine whether a large abundance of CO and CO$_2$, and corresponding small mixing ratio of CH$_4$, are also consistent with this dataset. For this purpose, we complete our {\it Spitzer} transit depths with those from EPOXI \citep[][0.35-1.0 $\mu$m]{Ballard2010TT},  HST WFC3 \citep[][1.14-1.65 $\mu$m]{Knutson2014}, HST NICMOS \citep[][1.1-1.9 $\mu$m]{Pont2009}, and from the ground in the {\it H} and {\it K} bands \citep{Alonso2008,Caceres2009} on Fig.~\ref{transit}. Here we model the planetary transit radius, which is inferred from the transit depth and \cite{vonBraun2012} stellar radius (0.455 R$_{\odot}$), as a function of wavelength, using the transmission spectrum code described in \citet{Fortney03,Fortney10b}.  The transmission spectrum of the planet GJ\,436b was previously modelled in some detail with this code by \citet{Shabram11}.  An advantage of these calculations is that we are able to model arbitrary chemical mixing ratios, rather than equilibrium chemistry.

We first compute the transmission spectrum of the 50$\times$-solar model, shown on Fig.~\ref{transit} in orange.  We compare to the two best-fit chemical models of \citetalias{Stevenson2010} in red and blue, which use low CH$_4$ and high CO and CO$_2$ mixing ratios.  The models differ dramatically in the optical only because it is unclear if the neutral atomic alkalis Na and K are still present in the planetary atmosphere, or have condensed into clouds.  The equilibrium chemistry model is strongly disfavoured by the \emph{Spitzer} data, as it yields the wrong planetary radius ratio between the 3.6 and 4.5 $\mu$m bands.  The blue and red curves reproduce the \emph{Spitzer} data better.  This is consistent with the dayside occultation data, where models with strongly enhanced CO and CO$_2$ and strongly depleted CH$_4$ are preferred.

We also note that deviations from a constant radius model are not particularly significant.  This could suggest that cloud material obscures the transmission spectrum \citep[e.g.,][]{Fortney05c}.  Alternatively, the atmospheric mean molecular weight could be higher that assumed.  A very metal-rich atmosphere would shrink the scale height as well as produce abundant `metal-metal' species such as CO and CO$_2$, which is certainly needed to explain the dayside spectrum.

\begin{figure}
\centering
  \includegraphics[width=9cm,trim=0 0 0 0,clip=true]{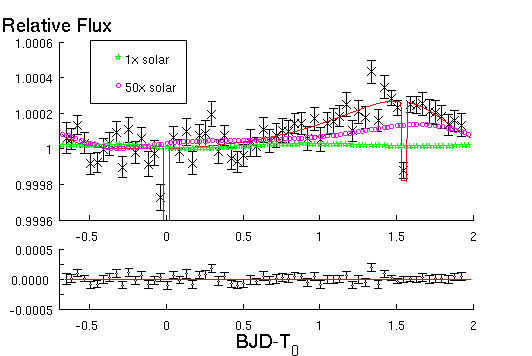}
  \caption{\label{phasei4}  \textbf{Top} : the observed phase curve of GJ\,436 at 8 $\mu$m (2008 July 12, AORs: 27863296, 27863552, 27863808) superimposed to a model for a Lambert sphere (red line). The adopted baseline models are respectively $p($[$xy$]$^{1}+w_{x}^{1}+w_{y}^{1}+l^{2})$, $p($[$xy$]$^{1}+w_{x}^{2}+w_{y}^{1})$, and $p($[$xy$]$^{1}+w_{x}^{1}+w_{y}^{1})$ to exclude temporal models. The black crosses are binned data with their 1-$\sigma$ error bars. The peak-to-peak flux difference equals 268 ppm while $A =  189 \pm 100 $ ppm.   Our model for a Lambert sphere is compared with \cite{Lewis2010} models shifted vertically.  The models correspond to 1x-solar (green stars) and 50x-solar (magenta circles) metallicity cases. The observed day-night temperature contrast indicates a  metal-rich atmosphere. \textbf{Bottom} : the residual light curve.}
\end{figure}

Our phase curve fit (Fig. \ref{phasei4}) presents a phase offset of ${-10^{  -  20 }_{  +  27} }^{\circ}$, a day-night contrast parameter $A =  189 ^{  -  91 }_{  +  100} $ ppm, which is not significant at a 3-$\sigma$ level,  and a peak-to-peak flux difference of 268~ppm. Those latter two values should be identical for a circular orbit model being perpendicular to the sky plane. We compare our phase curve fit with the 3D coupled radiative transfer and general circulation model   adapted to GJ\,436b by \citet{Lewis2010} (Fig.~\ref{phasei4}). It takes into account the pseudo-synchronous rotation of GJ\,436b and the atmosphere metallicity. The 1x-solar  atmospheric metallicity case of GJ\,436b (represented by green stars) does not fit the observations while the 50x-solar case (magenta circles) gets closer to them. If revealed significant, the amplitude disparity between the 3 light curves would require a (very) metal-rich atmosphere (\textgreater 50x-solar)  to explain the high day/night temperature contrasts.  Indeed, at higher metallicity, photons are absorbed and emitted from lower pressures, where the radiative timescale is short, making temperature homogenisation more difficult. This result would be in good qualitative agreement with our transmission and emission spectra analysis. 

\section{Discussion}
\label{sec:Disc}

\subsection{On the host star}
We could not find any hint of stellar activity (large amplitude trend, convincing flare-like structure) in our photometric time-series for \object{GJ\,436}. A low stellar activity for GJ\,436 is supported by the stable stellar flux (Fig.~\ref{fluxotime}), and by the temporal stability of the transit depths. In our light curves, we detect no occultation of stellar spots by the planet during its transits. Neither do \cite{Knutson2014}  in their HST WFC3 data. This is in good agreement with the advanced age of the star deduced from its kinematics \citep{Leggett1992} and weak chromospheric Ca {\scshape ii} H and K emission lines \citep{Butler2004}.

\begin{figure}
\centering
  \includegraphics[width=8.8cm,trim=0 0 10 10,clip=true]{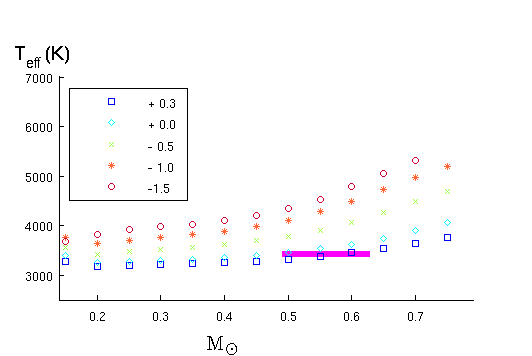}
  \caption{\label{modspada}  Theoretical effective temperature-mass relations from the stellar evolution model of \cite{Spada2013}, with the mixing length parameter $\alpha$ set to 0.5. Theoretical 4 Gyr isochrones are plotted for different metallicities indicated by legend.  The magenta filled box shows the GJ\,436 position on the diagram according to \cite{vonBraun2012}'s stellar temperature and our inferred stellar mass.}
\end{figure}

Our global analysis gives a stellar mass of $0.556 ^{+ 0.071}_{-0.065}$~M$_{\odot}$. This value is higher than what \cite{Delfosse2000} obtain from mass-luminosity relations partly calibrated with eclipsing binaries (0.44 $\pm$ 0.04~M$_{\odot}$). Our error on the stellar mass strongly depends on the stellar radius error. Multiplying by a factor two the stellar radius error almost doubles our uncertainty on the stellar mass ($\Delta M_{*}$=0.141~M$_{\odot}$). The fact that GJ\,436 appears as a M2.5 star despite its somewhat high mass might reflect an effect of the metallicity. Indeed, the stellar spectral type depends both on the mass and metallicity : for a given mass, the more metal-rich a star is, the lower its effective temperature is. For instance, GJ\,436 mass and temperature agree with \cite{Spada2013} models for a solar or supersolar metallicity (Fig. \ref{modspada}) as supported by our derived metallicity, and by e.g. \cite{Johnson2009} and \cite{Neves2013} with  [Fe/H] = +0.25 and +0.02 respectively.

\subsection{On the atmospheric model}

Our findings fit in reasonably well with recent theoretical advances.  Motivated by previous observational work on GJ\,436b, there have been recent suggestions along multiple theoretical lines that Neptune-class exoplanet atmospheres may commonly have extremely high metallicities, perhaps several hundred times solar, or higher.  \citet{Fortney13} have suggested, based on atmospheric accretion of planetesimals in population synthesis formation models, that high atmospheric metallicities may be a common outcome of the planet formation process.  \citet{Moses2013}, in a detailed chemical study of GJ\,436b, suggest that low CH$_4$ abundances, and high CO and CO$_2$ abundance can be a natural outcome in an atmosphere where metals are so abundant that H$_2$ is no longer the dominant atmospheric constituent, by number. Independently,  \citet{Agundez2013} favour models with an efficient tidal heating and also a high metallicity.  Finally, the featureless transmission spectrum extracted by \citet{Knutson2014} leads to an atmospheric metallicity of 1,900 times solar or to a cloud layer at pressure below 10 mbar.  It is not clear if such high envelope metallicity is consistent with the radius and bulk density of the planet.  We suggest continuing chemical studies of the atmosphere of GJ\,436b with our revised occultation and transit depths. It may possibly favour a high cloud seen in transmission without such a high atmospheric metallicity. Future comparisons with GJ\,3470b \citep{Bonfils2012,Crossfield2013,Venot2013,Demory2013} will also be illuminating.

\section{Conclusion}

We performed an independent and global analysis of all available {\it Spitzer} data for GJ\,436 combined with our new HARPS RV measurements. In this analysis, we optimised the data reduction procedure for each  {\it Spitzer} instrument with the adaptation of partial deconvolution photometry or aperture photometry, and we paid a particular attention to the modelling of the systematic effects. We recommend the use of the FWHM of the PSF in both directions as parameters to model the instrumental systematics. The insertion of the HARPS RVs complements well the photometric data.  Our results are globally consistent with previous studies \citepalias[e.g.,][]{Stevenson2010,Beaulieu2011,Knutson2011,Stevenson2012} but some discrepancies disclose another facet of GJ\,436b. In particular we obtained constant values of the transit depth with time, a flat transmission spectrum and a significantly lower 3.6~$\mu$m emission.

GJ\,436b is a warm Neptune with a mass of 25.4 $\pm$ 2.0 M$_{\oplus}$ and a radius of 4.096 $\pm$ 0.162 R$_{\oplus}$. It is in an eccentric orbit ($e = 0.162 \pm 0.004$) around a M2.5 star ($0.56 \pm0.06$ M$_{\odot}$). We detect no stellar variability (no stellar flux variation on a large scale and no transit depth variation with time) in the whole set of {\it Spitzer} light curves. No occulted star spots were observed in the transit light curves. \citetalias{Knutson2011} observed very weak photometric activity in the optical. In addition, Astudillo et al. (in prep) measure a S$_{HK}$ index in the HARPS spectra that is consistent with a weak stellar activity and does not show a detectable periodicity.  We thus confirm GJ\,436's weak activity. 

Neither the amplitude of the phase curve nor its shape can be constrained with the current data set. We recommend new observations with future facilities at multiple wavelengths longer than 8~$\mu$m combined with stellar monitoring  at shorter wavelengths, such as 4.5~$\mu$m or in the visible to discern the planetary phase curve from stellar variability. Such observations should constrain longitudinal properties of the atmosphere at different depths.  

Despite our shallower occultation depths at 3.6, 4.5, and 8~$\mu$m compared to previous works, all the photometric {\it Spitzer} time-series are still in good agreement with a metal-rich atmosphere depleted in methane and enhanced in CO/CO$_2$. However the metallicity of the atmosphere may not be as high than previously thought. A cooler atmospheric model with disequilibrium chemical abundance profiles should better fit our data, and also yield a lower flux in the 4.5~$\mu$m band, which has always been problematic to fit. We encourage an entirely new modelling analysis based on our revised data for firm conclusions on the joint constraint from the emission and transmission spectra.

We found no significant evidence for a second planet and constrained a maximum mass for a potential companion of 10 Earth masses up to few-hundred days period and of 3-5 Earth masses in GJ\,436's habitable zone.

%

\begin{acknowledgements}
This work is based in part on observations made with the Spitzer Space Telescope, which is operated by the Jet Propulsion Laboratory, California Institute of Technology under a contract with NASA. Support for this work was provided by NASA through an award issued by JPL/Caltech. M. Gillon is FNRS Research Associate. N. Astudillo acknowledges support from``Becas de Doctorado en el Extranjero, Becas Chile'' (grant 72120460). V. Neves and N. C. Santos acknowledge the support by the European Research Council/European Community under the FP7 through Starting Grant agreement number 239953. NCS was also supported by FCT through the Investigador FCT contract reference IF/00169/2012 and POPH/FSE (EC) by FEDER funding through the program``Programa Operacional de Factores de Competitividade - COMPETE''. We thank the anonymous referee for his helpful comments and suggestions as well as Pierre Drossart for helpful discussion about the methane fluorescence, Gwena\"el Bou\'e for information about ARoME use, Michael Line for providing his models from \citet{Moses2013} and Heather Knutson. We are grateful to Josefina Montalban for highlights on the stellar parameters and Thierry Morel for discussions about errors on inferred parameters for low mass stars. We thank Eva Eulaers, Alice Decock, and Sebastien Salmon for their helpful comments on the redaction. A special thank to Sandrine Sohy for her help and commitment in programming part of this work. 
\end{acknowledgements}

\bibliographystyle{aa} 
\bibliography{bibliogj} 

\end{document}